\documentclass[11pt]{revtex4}
\usepackage{amsmath}
\usepackage{amsfonts}
\usepackage{amssymb}
\usepackage{amsxtra}
\usepackage{supertabular}
\usepackage[pdftex]{graphicx}
\begin{document}
\title{Understanding the second quantization of fermions in Clifford and in Grassmann space\\
New way of second quantization of fermions\\
Part II}
\author{N.S. Manko\v c Bor\v stnik${}^1$ and H.B.F. Nielsen${}^2$
\\
${}^1$
University of Ljubljana, Slovenia\\
${}^2$Niels Bohr Institute, Denmark
}
%

\begin{abstract}
We present in Part II the description of the internal degrees of freedom of 
fermions by the superposition of odd products of the Clifford algebra elements, 
either $\gamma^a$'s or  $\tilde{\gamma}^a$'s~\cite{norma92,norma93,nh2018},
which determine with their oddness the anticommuting properties of the creation and annihilation operators  of the second quantized  fermion fields in even $d$-dimensional space-time, as we do in Part I of this paper by the Grassmann 
algebra elements $\theta^a$'s and  $\frac{\partial}{\partial \theta_a}$'s. 
We discuss: 
{\bf i.} The properties of the two kinds of the odd Clifford algebras, 
forming two independent spaces, both expressible with the Grassmann 
algebra of $\theta^{a}$'s and $\frac{\partial}{\partial \theta_{a}}$'s~\cite{norma93,nh02,nh03}.
{\bf ii.} The freezing out procedure of one of the two kinds of the odd Clifford 
objects, enabling that the remaining Clifford objects determine  with their oddness
in the tensor products of the finite number of the Clifford basis vectors and the 
infinite number of momentum basis, 
the creation and annihilation operators  carrying the family quantum numbers 
and fulfilling the anticommutation relations  of the second quantized fermions:
on the vacuum state, and on the whole Hilbert space defined by the sum of 
infinite number of "Slater determinants" of empty and occupied single fermion 
states.  {\bf iii.} The relation between the second quantized fermions as 
postulated by Dirac~\cite{Dirac,BetheJackiw,Weinberg} and the ones following 
from our Clifford algebra creation and annihilation operators, what offers the 
explanation for the Dirac postulates.

\end{abstract}



\maketitle

\section{Introduction}
\label{introduction}

In a long series of works we, mainly one of us N.S.M.B.~(\cite{norma92,%
norma93,IARD2016,n2014matterantimatter,nd2017,n2012scalars,%
JMP2013,normaJMP2015,nh2018} and the references therein), have found 
phenomenological success with the model named by N.S.M.B the 
{\it spin-charge-family}  theory,  with fermions, the internal space of which is describable as superposition of odd products of the Clifford algebra elements
$\gamma^a$'s in $d=(13+1)$ (may be  with 
d infinity), interacting with only gravity. The spins of fermions from higher 
dimensions, $d>(3+1)$, manifest in $d=(3+1)$ as charges of the 
{\it standard model}, the gravity originating in higher dimensions manifest
as the {\it standard model} vector gauge fields and the scalar Higgs explaining
the Yukawa couplings.

There are two kinds of anticommuting algebras, the Grassmann algebra and 
the Clifford algebra, the later with two independent subalgebras.
The Grassmann algebra, with elements $\theta^a$, and their Hermitian 
conjugated partners $\frac{\partial}{\partial \theta^a}$~\cite{nh2018},
describes fermions with the integer spins and charges in the adjoint 
representations, the two Clifford algebras, we call their elements 
$\gamma^a$ and $\tilde{\gamma}^a$, can each of them be used to 
describe half integer spins and charges in the fundamental representations. 
The Grassmann algebra  is expressible with the two Clifford algebras and 
opposite.

The two papers explain how do the oddness of the internal space of fermions manifests in the single particle wave functions, relating the oddness of the 
wave functions to the corresponding creation and annihilation operators of the 
to the second quantized fermions, in the Grassmann case and in the Clifford case,
explaining therefore the postulates of Dirac for the  second quantized fermions.

We learn in Part I of this paper, that in $d$-dimensional space $2^{d-1}$ superposition
of  odd products of $d$  $\,\,\theta^a$'s exist, 
chosen to be the eigenvectors of the Cartan subalgebra, Eq.~(4) of Part I, 
and arranged in tensor products with the momentum space to be solutions of the equation of motion for free massless ``fermions'', Eq.~(21) of Part I. 

The creation operators, defined as the tensor products of the superposition
of the finite number of ''basis vectors'' in Grassmann space, guaranteeing the 
oddness of operators, and of the infinite basis 
in momentum space, 
form --- applied on the vacuum state --- the second quantized states of integer spin ''Grassmann fermions''.  The creation operators 
fulfill together with their Hermitian conjugated partners annihilation operators 
(based on the internal space of odd products of 
$\frac{\partial}{\partial \theta_a}$'s) all the requirements of the 
anticommutation relations postulated by Dirac for fermions: {\bf i.} on the simple vacuum state $|\,1>$  (Eqs.~(7,11) of Part I), 
{\bf ii.} on the Hilbert space ${\cal H}$ ($= \prod_{\vec{p}}^{\infty}
\otimes_{N} {\cal H}_{\vec{p}}$, with the number of empty and occupied 
single fermion states for particular ${\vec{p}}$ equal to $2^{2^{d-1}}$)  of infinite 
many ''Slater determinants'' of all possible empty and occupied single fermion states (with the infinite number of possibilities of moments for each of   $2^{d-1}$ internal degrees of freedom), Eqs.~(25, 34) of Part I.
%

While the  creation and annihilation operators, which are superposition of odd products of $\theta^a$'s and $\frac{\partial}{\partial \theta_a}$'s, respectively, anticommute on the vacuum state $|\phi_o>=|\,1>$, Eq.~(7,11), the superposition of even products of $\theta^a$'s and $\frac{\partial}{\partial \theta_a}$'s, respectively, commute, Eq.~(16) of Part I. 

The superposition of odd products of $\gamma^a$'s and their Hermitian 
conjugated partners, as well as of odd products of $\tilde{\gamma}^a$'s 
and their Hermitian conjugated partners, on the corresponding vacuum states, Eq.~(\ref{vac1}),  anticommute. Since the tensor products of the ''basis vectors'' determining the internal space of Clifford fermions and of the basis in momentum space manifest oddness of the internal space,  no postulates of anticommutation relations as in the Dirac second quantization proposal is needed 
also  for Clifford fermions with the internal space described by one of the  two 
Clifford objects  (in Subsect.~\ref{postulates} we make a choice of 
$\gamma^a$'s).  The oddness of the '' basis vectors'', defining the internal space 
of fermions, transfers to the creation and annihilation operators  forming the second 
quantized single fermion states in the Clifford and the Grassmann space.


The  "Grassmann fermions" have  integer spins, and spins in the part with
$d\ge 5$  manifesting as charges in $d=(3+1)$, in adjoint representations, 
Table I in Part I. 
There is  no operator which would connect different irreducible representations of the corresponding Lorentz group. There are no elementary fermions with integer spin 
observed so far either.

The Clifford fermions, describing the internal space with $\gamma^a$'s, have half
integer spins and spins in the part with $d\ge 5$ manifesting as charges in $d=(3+1)$ in fundamental representations~\cite{normaJMP2015,n2014matterantimatter,%
IARD2016,nd2017,nh2017,nSin2018}. The operators $\tilde{S}^{ab}$ 
$(=\frac{i}{4} \{\tilde{\gamma}^a  \tilde{\gamma}^b - \tilde{\gamma}^b
\tilde{\gamma}^a)\}_{-})$ connect, after the reduction of the Clifford algebra 
degrees of freedom by a factor of $2$, Subsect.~\ref{postulates}, different irreducible representations of the Lorentz group $S^{ab}$ 
$(=\frac{i}{4} \{\gamma^a  \gamma^b - \gamma^b \gamma^a\}_{-})$
and determine ``family'' quantum numbers. All in agreement with the observed families of quarks and leptons.

In Part II the properties of the two kinds of the Clifford algebras objects, 
$\gamma^{a}$'s and $\tilde{\gamma}^{a}$'s, are discussed. Both are expressible
with $\theta^{a}$'s and $\frac{\partial}{\partial \theta_{a}}$'s
($\gamma^{a}= (\theta^{a} + \frac{\partial}{\partial \theta_{a}})$, 
$\tilde{\gamma}^{a}= i \,(\theta^{a} - 
\frac{\partial}{\partial \theta_{a}})$~\cite{norma93,nh02,nh03}),  
and both are, up to  a constant $\eta^{aa}=(1,-1,-1,\dots,-1)$,
Hermitian operators. Each of these two kinds of the Clifford algebra objects of an 
odd Clifford character (superposition of odd number of products of either
$\gamma^{a}$'s or $\tilde{\gamma}^{a}$'s, respectively) has $2^{d-1}$ members, 
together again $2\cdot 2^{d-1}$ members, the same as in the case of 
''Grassmann fermions''.

 These  two internal spaces, described by the two Clifford algebras, are 
 independent, each of them with their own
 generators of the Lorentz transformations, Eq.~(\ref{sabtildesab}), and their 
 corresponding   Cartan subalgebras, Eq.~(\ref{cartancliff}).

In each of these two internal spaces there exist $2^{\frac{d}{2}-1}$ 
''basis vectors'' in $2^{\frac{d}{2}-1}$ irreducible representations, chosen to 
be `'eigenvectors'' of  the 
corresponding Cartan subalgebra elements, Eq.~(\ref{eigencliffcartan}), and having the 
properties of creation and  annihilation operators (the Hermitian conjugated partners of
the creation operators) 
on the vacuum state: {\bf i.} The application  of any  creation operator on the vacuum 
state, Eq.~(\ref{vac1}), gives nonzero contribution, while the application of any 
annihilation operator on the vacuum state gives zero contribution. 
{\bf ii.} Within each of these two spaces 
all the annihilation operators anticommute among themselves  and all the creation 
operators anticommute among themselves. {\bf iii.} The vacuum state is a 
superposition of products of the annihilation operators with 
 their Hermitian conjugated partners creation operators, like in the Grassmann case.
The Clifford vacuum states, Eq.~(\ref{vac1}), are not the identity like in the Grassmann 
case, Eq.~(19) in Part~I. 

However,  there is not only the anticommutator of the creation operator and its 
Hermitian conjugated partner, which gives the nonzero contribution on the vacuum 
state in each of the two spaces --- what in the Grassmann algebra is the case, and
what the postulates of Dirac require. There are, namely, the additional 
($2^{\frac{d}{2}-1}-1$) members of the same irreducible representation, to which the 
Hermitian conjugated partner of the creation operator belongs, giving the nonzero 
anticommutator  with this creation operator on the vacuum state (Eq.~(\ref{should}) in 
Subsect.~\ref{propertiesCliffodd} illustrates such a case).

And, there is no operators, which would connect different irreducible representations 
in each of the two Clifford algebras and correspondingly there is no ``family'' quantum 
number for each irreducible representation, needed to describe the observed quarks 
and leptons. (Let the reader be reminded that also the Grassmann algebra has no 
operators, which would connect different irreducible representations. The Dirac's
second quantization postulates do not take care of charges and families of fermions, 
both can be treated and incorporated into the second quantization postulates as
quantum numbers of additional groups as proposed by the {\it standard model}.)
We solve these problems with the requirement, presented 
in Eq.~(\ref{tildegammareduced}): 
$\tilde{\gamma}^a B =(-)^B\, i \, B \gamma^a$,
with $(-)^B = -1$, if $B$ is (a function of) an odd product of $\gamma^a$'s,
 otherwise $(-)^B = 1$~\cite{nh03}.


We present in the subsection~\ref{steps} of this section a short overview
of steps, which lead to the second quantized fermions in the Clifford space,
offering the explanation for the Dirac's postulates. In the 
subsection~\ref{HNsubsection} we discuss our assumption, that the 
oddness of the ''basis vectors'' in the internal space transfer to the corresponding
creation and annihilation operators determining the second quantized single
fermion states and correspondingly the Hilbert space of the second quantized 
fermions, in a generalized way.

We present in Sect.~\ref{propertiesCliff0} the properties of the Clifford algebra 
"basis vectors" in the space of $d$ $\gamma^{a}$'s and in the space of $d$ 
$\tilde{\gamma}^{a}$'s.  
In Subsect.~\ref{propertiesCliffodd} we discuss properties of the ''basis vectors''
of half integer spin.
In Subsect.~\ref{postulates} we discuss conditions, under which operators  of  
one of these two kinds of the Clifford algebra objects demonstrate by themselves 
the anticommutation relations required for the second quantized "fermions", 
manifesting the half integer spins, offering the explanation for the spin and charges 
of the observed quarks and leptons and anti-quarks and anti-leptons and also for 
their families~%
\cite{norma92,norma93,IARD2016,n2014matterantimatter,nd2017,n2012scalars,%
JMP2013,normaJMP2015,nh2017,nh2018}. 

In Subsect.~\ref{Clifffamilies} we generate the basis states manifesting the family 
quantum numbers. 

In Subsect.~\ref{action} the superposition 
of "basis vectors", solving  the Weyl equation, are  constructed, forming   creation 
operators depending on the momenta and fulfilling with their Hermitian conjugated 
partners the anticommutation relations for the second quantized fermions.
 
We illustrate in Sect.~\ref{illustration} properties of the  Clifford odd 
''basis vectors'' in $d=(5+1)$-dimensional space, and extending  the internal space
in a tensor product to momentum space, we present also the superposition solving 
the Weyl equation, and correspondingly present creation and annihilation 
operators depending on the momentum $\vec{p}$.

We present in Sect.~\ref{HilbertCliff0} the Hilbert space ${\cal H}_{\vec{p}}$
of particular momentum $\vec{p}$ as "Slater determinants": i. with no "fermions" 
occupying any of the $2^{d-2}$ fermion states,  
ii. with one "fermion" occupying one of the $2^{d-2}$ fermion states, iii. with two 
"fermions" occupying the $2^{d-2}$ fermion states,..., up to the "Slater determinant"
 with all possible  fermion states of a particular  $\vec{p}$ occupied by "fermions".
The total  Hilbert space ${\cal H}$ is then the tensor product $\prod_{\infty}\otimes_{N}$ 
of infinite number of ${\cal H}_{\vec{p}}$. On ${\cal H}$ the tensor products 
of creation and annihilation operators (solving the equations of motion for free massless fermions) manifest the anticommutation relations of second quantized  "fermions" without any postulates. We also illustrate the application of the 
tensor products of creation  and annihillation operators on ${\cal H}$ in a
 simple toy model.

In Subsect.~\ref{bethenormarelation} the correspondence between our way and 
the Dirac way of second quantized fermions is presented, demonstrating that our
way does explain the Dirac's postulates.

In Sect.~\ref{13+1} we note that the present work is the part of the 
project named the {\it spin-charge-family} theory of one of the two authors of this
paper (N.S.M.B.).

In Sect.~\ref{conclusionsCliff} we comment on what we have learned from the 
second quantized integer spins "fermions", with the internal degrees of freedom 
described with Grassmann  algebra,  manifesting  (from the point of 
view of $d=(3+1)$) charges in the adjoint representations and compare these 
recognitions with the recognitions, which the Clifford algebra is offering for the 
description of fermions, 
appearing in families of the irreducible representations of the Lorentz group 
in the internal --- Clifford --- space, with half integer spins and charges and family
quantum numbers in the fundamental 
representations~\cite{norma92,norma93,IARD2016,n2014matterantimatter,nd2017,n2012scalars,%
JMP2013,normaJMP2015,nh2018}. 
%


%
\subsection{
Steps leading to second quantized Clifford fermions}
\label{steps}

We claim that when the internal part of the single particle wave functions
anticommute under the Clifford algebra product $*_A$, then  
the wave functions with such internal part, extended with a tensor product 
to momentum space,  anticommute  as well, and so do anticommute
the creation  and annihillation operators,  creating and annihilating the extended
fermion states, assuming  that the oddness of the algebra of the wave function extends to the creation and annihilation operators as presented in 
Subsect.~\ref{HNsubsection}.

If the internal part commute with respect to $*_A$ then the
corresponding wave functions and the creation operators commute as well.





Let us present steps which lead to the second quantized Clifford fermions, 
when using the odd Clifford algebra objects to define their internal space:\\ 
 {\bf i.} The superposition of an odd number of the Clifford algebra elements, either of $\gamma^{a}$'s or of $\tilde{\gamma}^a$, each with $2\cdot (2^{\frac{d}{2}-1})^2$ degrees of freedom, is used to describe the internal space of fermions in even dimensional spaces. \\
{\bf ii.} The "basis vectors" --- the superposition of an odd number of Clifford algebra elements --- are chosen to be the "eigenvectors" of the Cartan subalgebras, Eq.~(\ref{cartancliff}),  of the corresponding Lorentz algebras, 
Eq.~(\ref{sabtildesab}), in each of the two algebras.\\ 
{\bf iii.} There are two groups of $2^{\frac{d}{2}-1}$ members of 
$2^{\frac{d}{2}-1}$ irreducible representations of the corresponding Lorentz group, for either $\gamma^{a}$'s or for $\tilde{\gamma}^a$ algebras, each member of one group has its Hermitian conjugated partner in another group. \\ 
 Making a choice of one group of ''basis vectors'' (for either $\gamma^{a}$'s 
or for $\tilde{\gamma}^a$) to be creation operators, the other group of 
''basis vectors'' represents the annihilation operators. The creation operators 
anticommute among themselves and so do anticommute annihilation operators.\\ 
{\bf iv.} The vacuum state is then (for either  $\gamma^{a}$'s or for 
$\tilde{\gamma}^a$'s algebras) the superposition of products of annihilation 
$\times$ their Hermitian conjugated partners the creation operators.\\
 The application of the creation operators  on the  vacuum state 
 forms the "basis states" in each of the two spaces. The application of the annihilation operators on the vacuum state gives zero, Subsect.~\ref{HNsubsection}.\\ 
{\bf v.} The  requirement that application of $\tilde{\gamma}^a$ on $\gamma^a$ gives $- i \eta^{aa}$, and the application of $\tilde{\gamma}^a$ on identity  
gives $ i \eta^{aa}$ and that only $\gamma^a$'s are used to determine the 
internal space of half integer fermions, Eq.~(\ref{postulates}), reduces the 
dimension of the Clifford algebra for a factor of two, enabling that the Cartan subalgebra of $\tilde{S}^{ab}$'s  determines the ''family'' quantum numbers 
of each irreducible representation of $S^{ab}$'s, Eq.~(\ref{sabtildesab}), and 
correspondingly also of their Hermitian conjugated partners.\\
{\bf vi.} The tensor products of superposition of the finite number of members 
of the "basis vectors" and the infinite dimensional momentum basis, chosen to  
solve the Weyl equations for free massless half integer spin fermions, determine 
the creation and (their Hermitian conjugated partners) annihilation operators, 
which  depend  on the momenta $\vec{p}$, while $|p^0| =|\vec{p}|$ 
 ($p^a=(p^0,p^1, p^2, p^3, p^5,\dots, p^d)$), manifesting the  properties 
 of the observed fermions. 
These creation and annihilation operators fulfill on the Hilbert space all the 
requirements for the second quantized fermions, postulated by Dirac, 
 Eq.~(\ref{Weylpp'comrel})~\cite{Dirac,BetheJackiw,Weinberg}.\\
{\bf vii.} The second quantized Hilbert space ${\cal H}_{\vec{p}}$ of a particular
$\vec{p}$ is a tensor  product of creation operators of a particular $\vec{p}$, defining  "Slater determinants" with no single particle state occupied (with no creation operators applying on the vacuum state), with one single particle state occupied (with one creation operator applying on the vacuum state), with two single particle states occupied, and 
so on, defining in $d$-dimensional space $2^{(2^{\frac{d}{2}-1})^2}$ dimensional 
space for each $\vec{p}$.\\
{\bf viii.} Total Hilbert space is the infinite product  ($\otimes_{N}$) of 
${\cal H}_{\vec{p}}\,$: $\,{\cal H} $ 
 $= \prod_{\vec{p}}^{\infty}\otimes_{N} {\cal H}_{\vec{p}}$. 
The notation $\otimes_{N}$ is to point out that  odd algebraic products of the 
Clifford $\gamma^a$'s operators anticommute no matter for which  $\vec{p}$ 
they define  the orthonormalized superposition of 
''basis vectors'', solving the equations of motion as the orthonormalized plane 
wave solutions with $p^0 =|\vec{p}|$ and that the anticommutation character
keeps also in the tensor product of internal basis and momentum basis.\\
Since  the momentum space belonging to different $\vec{p}$ satisfy the "orthogonality" relations, the creation and annihilation operators
determined by $\vec{p}$ anticommute with the creation and annihilation operators determined by any other $\vec{p}{\,}'$. This means that in what ever way the Hilbert space ${\cal H}$ is arranged, the sign is changed 
whenever a creation or an annihilation operator, applying on the Hilbert space 
${\cal H}$, jumps over odd number of occupied states. 
No postulates for the second quantized fermions are needed in our odd Clifford 
space with creation and annihilation operators carrying the family quantum numbers.\\
{\bf x.} Correspondingly the creation and annihilation operators with the internal 
space described by either odd Clifford or odd Grassmann algebra, since fulfilling 
the anticommutation relations required for the second quantized fermions without 
postulates, explain the Dirac's postulates for the second quantized fermions.\\



%
%
\subsection{Our main assumption and definitions}
\label{HNsubsection}

(This subsection is the same as the one of Part I.)

In this subsection we clarify how does the main assumption of Part I and 
Part II: {\it the decision to describe the internal space of fermions with 
the ''basis vectors'' expressed with the superposition of odd products of 
the anticommuting members of the algebra}, either the Clifford one or 
the Grassmann one, acting algebraically, $*_{A}$, on the internal 
vacuum state  $|\psi_{o}>$, relate to the creation and annihilation 
anticommuting operators of the second quantized  fermion fields. 

To appreciate the need for this kind of assumption, 
let us first have in mind that algebra with the product $*_A$ is 
only present in our work, usually not in other works, and thus has no 
well known physical meaning. It is at first a product by which you can 
multiply two internal wave functions $B_{i}$ and $B_{j}$ with each other,
\begin{eqnarray}
\label{HNA}
C_k&=& B_i *_{A}B_j\,,\nonumber\\
B_i *_{A}B_j&=& \mp B_j *_{A} B_i \,, \nonumber 
\end{eqnarray} 
the sign $\mp$ depends on whether $B_i$ and $B_j$ are products of 
odd or even number of algebra elements: The sign is $-$ if both 
are (superposition of) odd products of algebra elements, in all other 
cases the sign is $+$.

Let ${\bf R}^{d-1}$ define the external spatial or momentum space. 
Then the tensor product $*_{T}$ extends the internal wave functions 
into the wave functions ${\bf C}_{\vec{p} ,\,i}$ defined in both spaces  
\begin{eqnarray}
\label{HNT}
{\bf C}_{\vec{p}, \, i} =|\vec{p}>*_{T}|B_{i}>\,,\nonumber
\end{eqnarray}
where again $B_i $ represent the superposition of products of elements 
of the  anticommuting algebras, in our case either 
$\theta^a$ or $\gamma^a$ or $\tilde{\gamma}^a$, 
used in this paper.

We can make a choice of the orthogonal and normalized basis so that 
$<{\bf C}_{\vec{p}, i} | {\bf C}_{\vec{p'}, j} > = 
\delta(\vec{p}  \vec{p'}) \,\delta_{ij}$. Let us point out that either 
$B_{i}$ or ${\bf C}_{\vec{p}, \, i}$ apply algebraically  on the vacuum 
state, $B_i*_{A} |\psi_o>$ and ${\bf C}_{\vec{p}, \, i}*_{A}|\psi_o>$.

Usually a product of single particle wave functions is not taken to have any 
physical meaning in as far as most physicists simply do not work with such 
products at all. 

To give to  the algebraic product, $*_A$,  and to the tensor product, $*_{T}$, 
defined on the space of single particle wave functions, the physical meaning, 
we postulate the connection  between the anticommuting/commuting properties 
of the ''basis  vectors'', expressed with  the odd/even products of the anticommuting 
algebra elements and the corresponding creation operators, creating second 
quantized single fermion/boson states
\begin{eqnarray}
\label{HNb}
{\hat b}^{\dagger}_{{\bf C}_{\vec{p}, i}} *_{A}|\psi_o> &=& |\psi_{\vec{p}, i}>\,,\nonumber\\
{\hat b}^{\dagger}_{{\bf C}_{\vec{p}, i}}  *_{T}\,|\psi_{\vec{p'}, j}> &=& 0\,,\nonumber\\
{\rm if \,} \vec{p} &=& \vec{p'}\, {\rm and} \, i=j\,,\nonumber\\
{\rm in\; all\;other\;cases\;} & &{\rm it\;follows\,}\nonumber\\
{\hat b}^{\dagger}_{{\bf C}_{\vec{p}, i}}  *_{T}\, 
{\hat b}^{\dagger}_{{\bf C}_{\vec{p'}, j}} *_{A}|\psi_o>
&=& \mp \,{\hat b}^{\dagger}_{{\bf C}_{\vec{p'}, j}}  *_{T} \, 
{\hat b}^{\dagger}_{{\bf C}_{\vec{p}, i}}  *_{A}|\psi_o>\,,\nonumber
\end{eqnarray} 
with the sign $\pm $ depending on whether  ${\hat b}^{\dagger}_{{\bf C}_{\vec{p}, i}}$
have both an odd character, the sign is $-$, or not, then the sign is $+$.

To each creation operator ${\hat b}^{\dagger}_{{\bf C}_{\vec{p}, i}}$ its Hermitian 
conjugated partner  represents the annihilation operator ${\hat b}_{{\bf C}_{\vec{p}, i}}$
\begin{eqnarray} 
\label{HNC}
{\hat b}_{{\bf C}_{\vec{p}, i}} &=&({\hat b}^{\dagger}_{{\bf C}_{\vec{p}, i}})^{\dagger}\,,
\nonumber\\
{\rm with \; the}&&{\rm property}\nonumber\\
{\hat b}_{{\bf C}_{\vec{p}, i}}\, *_{A}\,|\psi_o> &=&0\,,\nonumber\\
{\rm defining\; the\; } && {\rm vacuum \; state\; as\;}\nonumber\\
|\psi_o> :&=&\sum_i \,(B_{i})^{\dagger}\, *_{A} \,B_{i}
\,|\;I>\nonumber
\end{eqnarray}
where summation $i$ runs over all different products of annihilation operator 
$\times$ its Hermitian conjugated  creation operator, no matter for what 
$\vec{p}$ , and $|\;I>$ represents the identity, $(B_{i})^{\dagger}$ represents
the Hermitian conjugated wave function to $B_i$. 

Let the tensor multiplication $*_{T}$ denotes also  the multiplication of any number
of single particle states, and correspondingly of any number of creation operators.

What further means that to each single particle wave function we define 
the creation operator ${\hat b}^{\dagger}_{{\bf C}_{\vec{p}, i}} $, applying 
in a tensor product from the left hand side on the second quantized 
Hilbert space --- consisting of all possible products 
of any number of the single particle wave functions --- adding to the Hilbert 
space the single particle wave function created by this particular creation 
operator. In the case of the second quantized fermions, if this particular 
wave function with the quantum numbers and $\vec{p}$  of 
${\hat b}^{\dagger}_{{\bf C}_{\vec{p}, i}} $ is already 
among the single fermion wave functions of a particular product of 
fermion wave functions, the action of the creation operator gives zero, 
otherwise the number of the fermion wave functions increases for one. 
In the boson case the number of boson second quantized wave functions 
increases always  for one. 

If we apply with the annihilation operator ${\hat b}_{{\bf C}_{\vec{p}, i}} $ 
on the second quantized Hilbert space, then the application gives a 
nonzero contribution only if  the particular products of the single particle 
wave functions do include the wave function with the quantum number 
$i$ and $\vec{p}$. 

In a Slater determinant formalism the single particle wave functions define
the empty or occupied places of any of infinite numbers of Slater determinants. 
The creation operator ${\hat b}^{\dagger}_{{\bf C}_{\vec{p}, i}} $ applies
on a particular Slater determinant from the left hand side. Jumping over 
occupied states to the place with its $i$ and $\vec{p}$. If this state is 
occupied, the application gives in the fermion case zero, in the boson case 
the number of particles 
increase for one. The particular Slater determinant changes sign in the 
fermion case if 
${\hat b}^{\dagger}_{{\bf C}_{\vec{p}, i}}$ jumps over odd numbers of occupied 
states.  In the boson case the sign of the Slater determinant does not change. 

When annihilation operator ${\hat b}_{{\bf C}_{\vec{p}, i}} $ applies on 
particular Slater determinant, it is jumping over occupied states to its own place,
giving zero, if this space is empty and decreasing the number of occupied states,
if this space is occupied. The Slater determinant changes sign in the fermion 
case, if the number of occupied states before its own space is odd. In the boson
case the sign does not change.

Let us stress that choosing antisymmetry or symmetry is a choice which we make 
when treating  fermions or bosons, respectively, namely 
the choice of using oddness or evenness of basis vectors, that is the choice of using 
odd products or even products  of algebra anticummuting elements. 

To describe the second quantized fermion states  we make a choice of the basis 
vectors, which are the superposition of the odd numbers of algebra elements,
of both Clifford and Grassmann algebras.

The creation operators and their Hermitian conjugation partners annihilation 
operators therefore in the fermion case anticommute. The single fermion 
states, which are the application of the creation operators on the vacuum 
state $|\psi_o>$, manifest correspondingly as well the oddness. 
The vacuum state, defined as the sum over all different products of 
annihilation $\times$ the corresponding creation operators, have an even
character.

Let us end up with the recognition:\\ 
One usually means antisymmetry when talking about Slater-\underline{determinants}
because otherwise one would not get determinants.

In the present paper~\cite{norma92,norma93,IARD2016,nh02} the choice of 
the symmetrizing versus antisymmetrizing  relates indeed the commutation versus anticommutation with respect to the a priori completely different product $*_A$, 
of anticommuting members of the Clifford or Grassmann algebra. The oddness or 
evenness  of these products transfer to quantities to which these algebras
extend.

\section{Properties of Clifford algebra in even dimensional spaces}
\label{propertiesCliff0}

We  can learn in Part I that in $d$-dimensional space of anticommuting Grassmann 
coordinates (and of their Hermitian conjugated partners --- derivatives), Eqs.~(2,6) of Part I, 
there exist two kinds of the Clifford coordinates (operators) --- 
$\gamma^{a}$ and $\tilde{\gamma}^{a}$ --- both are expressible in terms of 
$\theta^{a}$ and their conjugate momenta $p^{\theta a}= i \,
\frac{\partial}{\partial \theta_{a}}$ ~\cite{norma93}.
\begin{eqnarray}
\label{clifftheta}
\gamma^{a} &=& (\theta^{a} + \frac{\partial}{\partial \theta_{a}})\,, \quad 
\tilde{\gamma}^{a} =i \,(\theta^{a} - \frac{\partial}{\partial \theta_{a}})\,,\nonumber\\
\theta^{a} &=&\frac{1}{2} \,(\gamma^{a} - i \tilde{\gamma}^{a})\,, \quad 
\frac{\partial}{\partial \theta_{a}}= \frac{1}{2} \,(\gamma^{a} + i \tilde{\gamma}^{a})\,,
\end{eqnarray}
offering together  $2\cdot 2^d$  operators: $2^d$ of those which are products of 
$\gamma^{a}$  and  $2^d$ of those which are products of $\tilde{\gamma}^{a}$.

Taking into account Eqs.~(1,2) of Part I ($\{\theta^{a}, \theta^{b}\}_{+}=0$,
$\{\frac{\partial}{\partial \theta_{a}}, \frac{\partial}{\partial \theta_{b}}\}_{+} =0$,
$\{\theta_{a},\frac{\partial}{\partial \theta_{b}}\}_{+} = \delta_{ab}$,
$\theta^{a \dagger} 
=\eta^{aa}\, \frac{\partial}{\partial \theta_{a}} $ and 
$(\frac{\partial}{\partial \theta_{a}})^{\dagger}=\eta^{aa} \theta^{a}$) one finds
\begin{eqnarray}
\label{gammatildeantiher}
\{\gamma^{a}, \gamma^{b}\}_{+}&=&2 \eta^{a b}= \{\tilde{\gamma}^{a}, 
\tilde{\gamma}^{b}\}_{+}\,, \nonumber\\
\{\gamma^{a}, \tilde{\gamma}^{b}\}_{+}&=&0\,,\quad
 (a,b)=(0,1,2,3,5,\cdots,d)\,, \nonumber\\
(\gamma^{a})^{\dagger} &=& \eta^{aa}\, \gamma^{a}\, , \quad 
(\tilde{\gamma}^{a})^{\dagger} =  \eta^{a a}\, \tilde{\gamma}^{a}\,,
\end{eqnarray}
with $\eta^{a b}=diag\{1,-1,-1,\cdots,-1\}$.

It follows  for the generators of the Lorentz algebra of each of the two kinds of the Clifford 
algebra operators, $S^{ab}$ and $\tilde{S}^{ab}$, that:
\begin{eqnarray}
S^{ab} &=&\frac{i}{4}(\gamma^a \gamma^b - \gamma^b \gamma^a)\,,\quad  
\tilde{S}^{ab} =\frac{i}{4}(\tilde{\gamma}^a \tilde{\gamma}^b - \tilde{\gamma}^b
 \tilde{\gamma}^a)\, , \nonumber\\
{\cal {\bf S}}^{ab} &=&S^{ab} + \tilde{S}^{ab}\,, \quad
\{S^{ab}, \tilde{S}^{ab}\}_{-}=0\,,\nonumber\\
\{ S^{ab}, \gamma^c\}_{-}&=& i (\eta^{bc} \gamma^a - \eta^{ac} \gamma^b)\,, \nonumber\\
\{ \tilde{S}^{ab}, \tilde{\gamma}^c \}_{-}&=& i (\eta^{bc} \tilde{\gamma}^a - 
\eta^{ac }\tilde{\gamma}^b)\,,\nonumber\\
\{ S^{ab}, \tilde{\gamma}^c\}_{-}&=&0\,,\quad \{\tilde{S}^{ab}, \gamma^c\}_{-}=0\,,
\label{sabtildesab}
\end{eqnarray}
where  ${\cal {\bf S}}^{ab}= i \, (\theta^{a} \frac{\partial}{\partial \theta_{b}} - \theta^{b}
\frac{\partial}{\partial \theta_{a}})$, 
Eq.~(3) of Part I.

Let us make a choice of the Cartan subalgebra of the commuting operators of
the Lorentz algebra for each of the two kinds of the operators of the Clifford algebra, $S^{ab}$ and $\tilde{S}^{ab}$, equivalent to the choice of 
Cartan subalgebra of ${\cal {\bf S}}^{ab}$ in the Grassmann case, Eq.~(4)
in Part I,
\begin{eqnarray}
S^{03}, S^{12}, S^{56}, \cdots, S^{d-1 \;d}\,, \nonumber\\
\tilde{S}^{03}, \tilde{S}^{12}, \tilde{S}^{56}, \cdots,  \tilde{S}^{d-1\; d}\,. 
\label{cartancliff}
\end{eqnarray}

Representations of $\gamma^a$ and representations of  $\tilde{\gamma}^a$ are 
independent, each with twice $2^{\frac{d}{2}-1}$ members in $2^{\frac{d}{2}-1}$
irreducible representations of an odd Clifford character and with  twice 
$2^{\frac{d}{2}-1}$ members in $2^{\frac{d}{2}-1}$irreducible representations of 
an even Clifford character in even dimensional spaces.

We make a choice for the members of the irreducible representations of the two Lorentz 
groups to be the  "eigenvectors" of the corresponding Cartan subalgebra of 
Eq.~(\ref{cartancliff}),  taking into account Eq.~(\ref{gammatildeantiher}),  
\begin{eqnarray}
S^{ab} \frac{1}{2} (\gamma^a + \frac{\eta^{aa}}{ik} \gamma^b) &=& \frac{k}{2}  \,
\frac{1}{2} (\gamma^a + \frac{\eta^{aa}}{ik} \gamma^b)\,,\nonumber\\
S^{ab} \frac{1}{2} (1 +  \frac{i}{k}  \gamma^a \gamma^b) &=&  \frac{k}{2}  \,
 \frac{1}{2} (1 +  \frac{i}{k}  \gamma^a \gamma^b)\,,\nonumber\\
\tilde{S}^{ab} \frac{1}{2} (\tilde{\gamma}^a + \frac{\eta^{aa}}{ik} \tilde{\gamma}^b) &=& 
\frac{k}{2}  \,\frac{1}{2} (\tilde{\gamma}^a + \frac{\eta^{aa}}{ik} \tilde{\gamma}^b)\,,
\nonumber\\
\tilde{S}^{ab} \frac{1}{2} (1 +  \frac{i}{k}  \tilde{\gamma}^a \tilde{\gamma}^b) &=& 
 \frac{k}{2}  \, \frac{1}{2} (1 +  \frac{i}{k} \tilde{\gamma}^a \tilde{\gamma}^b)\,.
\label{eigencliffcartan}
\end{eqnarray}
The Clifford "vectors" --- nilpotents and projectors --- of both algebras are normalized,
 up to a phase, with respect to Eq.~(\ref{grassintegral}) of \ref{normgrass}.
Both have half integer spins. The "eigenvalues" of the operator $S^{03}$, for example,  
for the "vector" $\frac{1}{2} (\gamma^0 \mp \gamma^3)$ are equal to 
$\pm\, \frac{i}{2}$, respectively, for the "vector" $\frac{1}{2} (1\pm \gamma^0  \gamma^3)$
are $\pm\, \frac{i}{2}$, respectively, while all the rest "vectors" have "eigenvalues" 
$\pm\, \frac{1}{2}$. One finds equivalently for the "eigenvectors" of the operator 
$\tilde{S}^{03}$: for $\frac{1}{2} \,( \tilde{\gamma^0} \mp \tilde{\gamma}^3)$  
the "eigenvalues"
$\pm\, \frac{i}{2}$, respectively, and for the "eigenvectors" $\frac{1}{2} 
(1\pm \tilde{\gamma}^0  \tilde{\gamma}^3)$ the "eigenvalues" $k=\pm\,  \frac{i}{2}$, 
respectively, while all the rest "vectors" have $k=\pm\,  \frac{1}{2}$. 

To make discussions easier let us introduce the notation for the "eigenvectors" of the 
two Cartan subalgebras, Eq.~(\ref{cartancliff}), Ref.~\cite{nh02,norma93}.
\begin{eqnarray}
\stackrel{ab}{(k)}:&=& 
\frac{1}{2}(\gamma^a + \frac{\eta^{aa}}{ik} \gamma^b)\,,\quad 
\stackrel{ab}{(k)}^{\dagger} = \eta^{aa}\stackrel{ab}{(-k)}\,,\quad 
(\stackrel{ab}{(k)})^2 =0\,,\nonumber\\
\stackrel{ab}{[k]}:&=&
\frac{1}{2}(1+ \frac{i}{k} \gamma^a \gamma^b)\,,\quad \;\,
\stackrel{ab}{[k]}^{\dagger} = \,\stackrel{ab}{[k]}\,, \quad \quad \quad \quad
(\stackrel{ab}{[k]})^2 = \stackrel{ab}{[k]}\,, \nonumber\\
\stackrel{ab}{\tilde{(k)}}:&=& 
\frac{1}{2}(\tilde{\gamma}^a + \frac{\eta^{aa}}{ik} \tilde{\gamma}^b)\,,\quad 
\stackrel{ab}{\tilde{(k)}}^{\dagger} = \eta^{aa}\stackrel{ab}{\tilde{(-k)}}\,,\quad
(\stackrel{ab}{\tilde{(k)}})^2=0\,,\nonumber\\
\stackrel{ab}{\tilde{[k]}}:&=&
\frac{1}{2}(1+ \frac{i}{k} \tilde{\gamma}^a \tilde{\gamma}^b)\,,\quad \;\,
\stackrel{ab}{\tilde{[k]}}^{\dagger} = \,\stackrel{ab}{\tilde{[k]}}\,,
\quad \quad \quad \quad
(\stackrel{ab}{\tilde{[k]}})^2=\stackrel{ab}{\tilde{[k]}}\,,\nonumber\\
\label{graficcliff}
\end{eqnarray}
with $k^2 = \eta^{aa} \eta^{bb}$. Let us notice that the ``eigenvectors'' of
 the Cartan subalgebras are either projectors 
\[ (\stackrel{ab}{[k]})^2= \stackrel{ab}{[k]}\,,\qquad (\stackrel{ab}{\tilde{[k]}})^2= 
\stackrel{ab}{\tilde{[k]}}\,,\]
or nilpotents 
\[(\stackrel{ab}{(k)})^{2}=0\,,\qquad (\stackrel{ab}{\tilde{(k)}})^{2}=0\,.\] 
We pay attention on even dimensional spaces, $d=2(2n+1)$ or $d=4n$, $n\ge0$.

The "basis vectors", which are products of $\frac{d}{2}$ either of nilpotents or
of projectors or of both, are ``eigenstates`' of all the members of the Cartan 
subalgebra, Eq.~(\ref{cartancliff}), of the corresponding Lorentz algebra, forming 
$2^{\frac{d}{2}-1}$ irreducible representations with $2^{\frac{d}{2}-1}$ members 
in each of the two  Clifford algebras cases. 

The "basis vectors" of Eq.~(\ref{allcartaneigenvec}) are "eigenvectors"  of all 
the Cartan subalgebra members, Eq.~(\ref{cartancliff}), in
 $d=2(2n+1)$-dimensional space of $\gamma^a$'s.
The first one is the product of  nilpotents only and correspondingly a superposition 
of an odd products of $\gamma^a$'s. The second one belongs to the same 
irreducible representation as the first one, if it follows from the first one by the 
application of $S^{01}$, for example.
%
\begin{eqnarray}
\label{allcartaneigenvec}
&&\stackrel{03}{(+i)}\stackrel{12}{(+)}\cdots \stackrel{d-1 \, d}{(+)}\,,\;\;\quad
 \stackrel{03}{[-i]}\stackrel{12}{[-i]} \stackrel{56}{(+)} \cdots 
\stackrel{d-1 \, d}{(+)}\,,  \nonumber\\ 
&&\stackrel{03}{[-i]} \stackrel{12}{[-]} \cdots \stackrel{d-1\,d}{[-]}\,.
\end{eqnarray}
One finds for their Hermitian conjugated partners, up to a sign,
\begin{eqnarray}
\label{allcartaneigenvecher}
&&\stackrel{03}{(-i)}\stackrel{12}{(-)}\cdots \stackrel{d-1 \, d}{(-)}\,,\;\;\quad
\ \stackrel{03}{[-i]}\stackrel{12}{[-i]} \stackrel{56}{(-)}\cdots 
\stackrel{d-1 \, d}{(-)}\,,\nonumber\\  
&& \stackrel{03}{[-i]} \stackrel{12}{[-]}\cdots \cdots\stackrel{d-1\,d}{[-]}\,.\
\nonumber
\end{eqnarray}

The "basis vectors" form an orthonormal basis within each of the irreducible 
representations or among irreducible representations, like the product of the 
following annihilation and the corres\-ponding creation operator:\\ 
$ \stackrel{d-1 \, d}{(-)}\cdots \stackrel{12}{(-)} \stackrel{03}{(-i)} *_{A}$
$\stackrel{03}{(+i)}\stackrel{12}{(+)}\cdots \stackrel{d-1 \, d}{(+)} =1$, while all the
algebraic products, which do not relate the annihilation operators with their Hermitian conjugated creation operators, give zero.  

Usually the operators $\gamma^a$'s are represented as matrices. We use $\gamma^a$'s
here to form the basis. One can find in Ref.~\cite{DMN} how does the application of
 $\gamma^a$'s on the basis defined in $d=(3+1)$ look like.

\subsection{Clifford ``basis vectors'' with half integer spin}
\label{propertiesCliffodd}

In the Grassmann case the $ 2^{d-1}$ odd and $ 2^{d-1}$ even Grassmann 
operators, which are superposition of either odd or even products of $\theta^a$'s, 
are well distinguishable from their $ 2^{d-1}$ 
odd and $ 2^{d-1}$ even Hermitian conjugated operators, which are superposition of  
odd and even  products of $\frac{\partial}{\partial \theta_{a}}$'s, Eq.~(6) in Part I.

In the Clifford case the relation 
between "basis vectors" and their Hermitian conjugated partners  (made 
of products of nilpotents ($\stackrel{ab}{(k)}$ or  $\stackrel{ab}{\tilde{(k)}}$) and 
projectors  ($\stackrel{ab}{[k]}$ or  $\stackrel{ab}{\tilde{[k]}}$), Eq.~(\ref{graficcliff}),
 are less transparent (although still easy to be evaluated).
%
%
 This can be noticed in Eq.~(\ref{graficcliff}), since 
$\frac{1}{\sqrt{2}} (\gamma^a + \frac{\eta^{aa}}{i\,k} \gamma^b)^{\dagger}$  is 
 $\eta^{aa}\,\frac{1}{\sqrt{2}} (\gamma^a + \frac{\eta^{aa}}{i \,(- k)} \gamma^b) $, 
while $ (\frac{1}{\sqrt{2}} (1 +  \frac{i}{k}  \gamma^a \gamma^b))^{\dagger}=$
$ \frac{1}{\sqrt{2}} (1 +  \frac{i}{k}  \gamma^a \gamma^b)$ is self adjoint. 
(This is the case also for representations in the sector of $\tilde{\gamma}^a$'s.)

One easily sees that in even dimensional spaces, either in $d=2(2n+1)$ or in $d=4n$, 
the Clifford odd "basis vectors" (they are products of an odd number of nilpotents and 
an even number of projectors) have their Hermitian conjugated partners in another
irreducible representation, since Hermitian conjugation changes an odd number of 
nilpotents (changing at the same time the handedness of the "basis vectors"), while the generators of the Lorentz transformations change two nilpotents at 
the same time (keeping the handedness unchanged).

The Clifford even "basis vectors" have an even number of nilpotents and can have 
an odd or an even number of projectors. Correspondingly an irreducible representation of an even "basis vector" can be a product of projectors only and therefore is self adjoint.

Let us recognize the properties of the nilpotents and projectors. The relations are taken 
from Ref.~\cite{IARD2016}.
\begin{eqnarray}
\stackrel{ab}{(k)}\stackrel{ab}{(k)}& =& 0\,, \quad \quad \stackrel{ab}{(k)}\stackrel{ab}{(-k)}
= \eta^{aa}  \stackrel{ab}{[k]}\,, \nonumber\\
\stackrel{ab}{[k]}\stackrel{ab}{[k]} &=& \stackrel{ab}{[k]}\,, \quad \quad
\stackrel{ab}{[k]}\stackrel{ab}{[-k]}= 0\,, \;\;\nonumber\\
\stackrel{ab}{(k)}\stackrel{ab}{[k]}& =& 0\,,\quad \quad \quad \stackrel{ab}{[k]}\stackrel{ab}{(k)}
=  \stackrel{ab}{(k)}\,, \nonumber\\
  \stackrel{ab}{(k)}\stackrel{ab}{[-k]} &=&  \stackrel{ab}{(k)}\,,
\quad \; \stackrel{ab}{[k]}\stackrel{ab}{(-k)} =0\,. 
\label{graphbinoms}
\end{eqnarray}
The same relations are valid also if one replaces $\stackrel{ab}{(k)} $ with 
$\stackrel{ab}{\tilde{(k)}}$ and  $\stackrel{ab}{[k]}$ with $\stackrel{ab}{\tilde{[k]}}$,
Eq.~(\ref{graficcliff}). 

Taking into account Eq.~(\ref{graphbinoms}) one recognizes that the product of 
annihilation and the creation operator from Eq.~(\ref{allcartaneigenvec}), 
$\stackrel{03}{(-i)}\stackrel{12}{(-)}\cdots \stackrel{d-1 \, d}{(-)}$ $*_{A}$
$\stackrel{03}{(+i)}\stackrel{12}{(+)}\cdots \stackrel{d-1 \, d}{(+)}$, applied 
on a vacuum state --- defined as a sum of products of all annihilation 
$\times$ their Hermitian conjugated partner creation operators from 
all irreducible representations,
$\stackrel{03}{[-i]}\stackrel{12}{[-]}\stackrel{56}{[-]}\cdots \stackrel{d-1 \, d}{[-]} +
\stackrel{03}{[+i]}\stackrel{12}{[+]}\stackrel{56}{[-]}\cdots \stackrel{d-1 \, d}{[-]} +
\stackrel{03}{[+i]}\stackrel{12}{[-]}\stackrel{56}{[+]} \stackrel{78}{[-]}\cdots
 \stackrel{d-1 \, d}{[-]} + \cdots$, Eq.~(\ref{vac1}),
gives a nonzero contribution, but is not the only one for a chosen creation operator. 
There are several other choices, like
\begin{eqnarray}
\label{notunique0} 
\stackrel{03}{[+i]}\stackrel{12}{[+]}\cdots \stackrel{d-1 \, d}{(-)} *_{A} 
\stackrel{03}{(+i)}\stackrel{12}{(+)}\cdots \stackrel{d-1 \, d}{(+)}\,,\nonumber\\
\stackrel{03}{[+i]}\stackrel{12}{(-)} \stackrel{56}{[+]}\cdots 
\stackrel{d-1 \, d}{(-)} *_{A} 
 \stackrel{03}{(+i)}\stackrel{12}{(+)}\cdots 
\stackrel{d-1 \, d}{(+)}\,,\nonumber
\end{eqnarray}
which also give nonzero contributions.

Let us recognize:

i. The two Clifford spaces, the one spanned by $\gamma^{a}$'s and the second one
spanned by $\tilde{\gamma}^{a}$'s, are independent vector spaces, each with 
$2^{d}$ "vectors". 

ii. The Clifford odd "vectors" (the superposition of products of 
odd numbers of $\gamma^a$'s or  $\tilde{\gamma}^{a}$'s, respectively)
can be arranged for each kind of the Clifford algebras into two groups of 
$2^{\frac{d}{2} - 1}$ members of $2^{\frac{d}{2} - 1} $ irreducible 
representations of the corresponding Lorentz group. The two groups are  
Hermitian conjugated to each other. 

iii. Different irreducible representations are indistinguishable with respect to the 
"eigenvalues" of the corresponding  Cartan subalgebra members.

iv. The Clifford even part (made of superposition of products of even numbers of  
$\gamma^a$'s and  $\tilde{\gamma}^{a}$'s, respectively) splits as well into 
twice $2^{\frac{d}{2} - 1} \cdot 2^{\frac{d}{2} - 1} $ irreducible representations 
of the Lorentz group. One member of each Clifford  even representation, the one
which is the product of projectors only, is self adjoint. Members of one irreducible 
representation are with respect to the Cartan subalgebra indistinguishable from
all the other irreducible representations.

%
v. The $2^{\frac{d}{2}-1}$ members of each of the $ 2^{\frac{d}{2}-1}$
irreducible representations are orthogonal to one another and so are orthogonal 
their corresponding Hermitian conjugated partners. For illustration of the 
orthogonality one can look at Table~\ref{Table Clifffourplet.},  and recognize 
that any ''basis vector'' of the first four multiplets of {\it odd I}, if multiplied from 
the left hand side or from the right hand side with any other ''basis vector'' from 
the rest three ''families'' of {\it odd I} get zero when taking into account Eq.~(\ref{graphbinoms}). One can repeat this also for any ''basis vectors'' of all
the ``families''  of {\it odd I}, as well as among all the ``basis vectors'' within
{\it odd II}. Generalization to any even dimension $d$ is straightforward.

vi. Denoting "basis vectors"  by $\hat{b}^{m \dagger}_{f}$,
(where $f$ defines different irreducible representations and $m$ a member in the 
representation $f$), 
and their Hermitian conjugate partners by  $\hat{b}^{m}_{f}$ 
$=(\hat{b}^{m \dagger}_{f})^{\dagger}$, let us start for $d=2(2n+1)$ with
\begin{eqnarray}
\label{bmfdagerbmf}
\hat{b}^{m=1 \dagger}_{f=1} \, {\bf :} &=&\stackrel{03}{(+i)}\stackrel{12}{(+)}
\cdots \stackrel{d-1 \, d}{(+)}\,,\nonumber\\
(\hat{b}^{m=1 \dagger}_{f=1})^{\dagger}=\hat{b}^{m=1 }_{f=1} \, {\bf :} &=& \, 
\stackrel{d-1 \, d}{(-)} \cdots \stackrel{12}{(-)}\stackrel{03}{(-i)}\,, 
\end{eqnarray}
 and making a choice of the vacuum state
 $|\psi_{oc}> $ as a sum of all the products of 
$\hat{b}^{m}_{f}\cdot \hat{b}^{m \dagger}_{f}$ for all 
$f=(1,2,\cdots,2^{\frac{d}{2}-1})$, one recognizes for the "basis vectors" of an 
odd Clifford character for each of the two Clifford algebras the properties
\begin{eqnarray}
\label{almostDirac}
\hat{b}^{m}_{f} {}_{*_{A}}|\psi_{oc}>&=& 0\, |\psi_{oc}>\,,\nonumber\\
\hat{b}^{m \dagger}_{f}{}_{*_{A}}|\psi_{oc}>&=&  |\psi^m_{f}>\,,\nonumber\\
\{\hat{b}^{m}_{f}, \hat{b}^{m'}_{f'}\}_{*_{A}+}|\psi_{oc}>&=&
 0\,|\psi_{oc}>\,, \nonumber\\
\{\hat{b}^{m \dagger}_{f}, \hat{b}^{m \dagger}_{f}\}_{*_{A}+}|\psi_{oc}>
&=&|\psi_{oc}>\,.
\end{eqnarray}
 ${*_{A}}$ represents the algebraic multiplication of $\hat{b}^{m \dagger}_{f}$ 
 and $ \hat{b}^{m'}_{f'} $  among themselves and  with the vacuum state 
 $|\psi_{oc}>$ of Eq.(\ref{vac1}), which takes into account 
Eq.~(\ref{gammatildeantiher}).
All the products of Clifford algebra elements are up to now the algebraic ones and so 
are also the products in Eq.~(\ref{almostDirac}). Since we use  here 
anticommutation relations, we pointed out with ${}_{*_{A}}$ this algebraic character
of the products, to be later distinguished from the tensor product ${}_{*_{T}}$,
when the creation and annihilattion operators  are defined on an extended basis, 
which is the tensor product of the superposition of the ''basis vectors'' of the Clifford 
space and of the momentum basis, applying on the Hilbert space of ''Slater determinants''. The tensor product $*_{T}$ is used as well as the product 
mapping a pair of the fermion wave functions in to two fermion wave functions 
and further to many fermion wave functions --- that is to the extended algebra 
of many fermion system.

Obviously, $\hat{b}^{m \dagger}_{f}$ and $\hat{b}^{m}_{f}$ have on the 
level of the algebraic products, when applying on the vacuum state $|\psi_{oc}>$,
{\it almost} the properties of creation and annihilation operators of the second 
quantized fermions in the postulates of Dirac,  as it is discussed in the next items. 
We illustrate properties of "basis vectors"  and their Hermitian conjugated 
partners on the example of $d=(5+1)$-dimensional space in 
Subsect.~\ref{illustration}.\\
 ${\quad}$ vii. a. There is, namely, the  property, which the  second quantized fermions should 
fulfill in addition to the relations of Eq.~(\ref{almostDirac}). The anticommutation 
relations of creation and annihilation operators should be:
\begin{eqnarray} 
\label{should}
\{\hat{b}^{m}_{f}, \hat{b}^{m'\dagger}_{f'}\}_{*_{A}+}|\psi_{oc}>&=&
\delta^{m m'} \delta_{f f'} |\psi_{oc}>\,.
\end{eqnarray}
For  any $\hat{b}^{m}_{f}$ and any $\hat{b}^{m'\dagger}_{f'}$ this is not 
the case;  besides $\hat{b}^{m=1}_{f=1} = \, 
\stackrel{d-1 \, d}{(-)} \cdots  \stackrel{56}{(-)} \stackrel{12}{(-)}\stackrel{03}{(-i)}$, for example, also 
\begin{eqnarray} 
\label{notunique1}
\hat{b}^{m'}_{f'} &=& \, 
\stackrel{d-1 \, d}{(-)} \cdots  \stackrel{56}{(-)}
 \stackrel{12}{[+]}\stackrel{03}{[+i]}\,,\nonumber
\end{eqnarray}
and several others give, when applied on $\hat{b}^{m=1\dagger}_{f=1}$,
nonzero contributions.
 There are namely $2^{\frac{d}{2}-1}-1$ too many annihilation
operators for each creation operator, which give, applied on the creation operator, 
nonzero contribution. \\
${\quad}$ vii. b. To use the Clifford algebra objects to describe second quantized fermions,
representing
the observed quarks and leptons as well as the antiquarks and antileptons%
~\cite{IARD2016,n2014matterantimatter,nd2017,%
n2012scalars,JMP2013,normaJMP2015,nh2017,nh2018},  {\it the families should exist}. \\
${\quad}$ vii. c. 
The operators should exist, which connect one irreducible 
representation of fermions with all the other irreducible representations.
\\
${\quad}$ vii. d. 
Two independent choices for describing the internal degrees 
of freedom of the observed quarks and leptons are not in agreement with the observed 
properties of fermions. 

We solve these problems, cited in vii. a., vii. b., vii. c.  and vii. d., by reducing  the degrees 
of freedom offered by  the two kinds of the Clifford algebras, $\gamma^a$'s
and  $\tilde{\gamma}^a$'s, making a choice of one --- $\gamma^a$'s --- to describe 
the internal space of fermions, and using the other one --- 
$\tilde{\gamma}^a$'s --- to describe the "family" quantum number of each irreducible
representation of $S^{ab}$'s in  space  defined by $\gamma^a$'s.



%
\subsection{Reduction of the Clifford space by the postulate}
\label{postulates}

The creation and annihilation operators of an odd Clifford  algebra of both kinds, of 
either $\gamma^a$'s or  $\tilde{\gamma}^{a}$'s,   would obviously 
obey the anticommutation relations for the second quantized fermions, 
postulated by Dirac, at least on the vacuum state, which is a sum of all the products 
of annihilation times, $*_{A}$, the corresponding creation operators, provided that 
each of the irreducible representations would carry a different quantum number.  

But we know that a particular member $m$ has for all the irreducible representations 
the same quantum numbers, that is the same "eigenvalues" of the 
Cartan subalgebra (for the vector space of either $\gamma^a$'s or  
$\tilde{\gamma}^{a}$'s), Eq.~(\ref{graficcliff}).  

{\it The only possibility to "dress" each irreducible representation of one kind of the
two independent vector spaces with a new, let us say   "family"  quantum number, 
is that we "sacrifice" one of the two vector spaces, let us make a choice of}  
$\tilde{\gamma}^{a}$'s,  {\it and use $\tilde{\gamma}^{a}$'s 
to define the "family" quantum number for each irreducible representation of the 
vector space of} $\gamma^a$'s, while  {\it  keeping the relations of} 
Eq.~(\ref{gammatildeantiher}) {\it  unchanged:}  $\{\gamma^{a}, 
\gamma^{b}\}_{+}=2 \eta^{a b}= \{\tilde{\gamma}^{a}, 
\tilde{\gamma}^{b}\}_{+}$, $\{\gamma^{a}, \tilde{\gamma}^{b}\}_{+}=0$,
  $ (\gamma^{a})^{\dagger} = \eta^{aa}\, \gamma^{a}$, 
$(\tilde{\gamma}^{a})^{\dagger} =  \eta^{a a}\, \tilde{\gamma}^{a}$, 
$(a,b)=(0,1,2,3,5,\cdots,d)$.\\

We therefore {\it postulate}:\\
  Let  $\tilde{\gamma}^{a}$'s operate on $\gamma^a$'s as follows~%
\cite{nh03,norma93,JMP2013,normaJMP2015,nh2018}

\begin{eqnarray}
\tilde{\gamma}^a B &=&(-)^B\, i \, B \gamma^a\,,
\label{tildegammareduced}
\end{eqnarray}

with $(-)^B = -1$, if $B$ is (a function of) an odd product of $\gamma^a$'s,
 otherwise $(-)^B = 1$~\cite{nh03}.\\ 

After this postulate the vector  space of $\tilde{\gamma}^{a}$'s is correspondingly 
 "frozen out". No vector space of $\tilde{\gamma}^{a}$'s 
needs to be taken into account  any longer, in agreement with the observed properties
of fermions. This solves the problems vii.a -  vii. d. of Subsect.~\ref{propertiesCliffodd}.\\

Taking into account Eq.~(\ref{tildegammareduced})  we can check that:\\
{\bf a.} Relations of Eq.~(\ref{gammatildeantiher}) remain unchanged~\footnote{
Let us show that the relation $ \{\tilde{\gamma}^{a}, 
\tilde{\gamma}^{b}\}_{+}= 2 \eta^{ab}$ remains valid when applied on $B$, if $B$ 
is either an odd or an even product of $\gamma^a$'s:
$ \{\tilde{\gamma}^{a}, \tilde{\gamma}^{b}\}_{+}$ $ \gamma^c=$
$-i \,(\tilde{\gamma}^{a} \gamma^c \gamma^b + \tilde{\gamma}^{b} \gamma^c \gamma^a)=$ $-i\,i\,\gamma^c ( \gamma^b \gamma^a +  \gamma^a \gamma^b)
 = 2 \eta^{ab} \gamma^c$, while $ \{\tilde{\gamma}^{a}, \tilde{\gamma}^{b}\}_{+}$ 
 $ \gamma^c \gamma^d=$ $ i \,(\tilde{\gamma}^{a} \gamma^c \gamma^d  \gamma^b + \tilde{\gamma}^{b} \gamma^c \gamma^d \gamma^a)=$ $i(-i)\,\gamma^c \gamma^d (\gamma^b \gamma^a +  \gamma^a \gamma^b)
 = 2 \eta^{ab} \gamma^c \gamma^d$. The relation is valid for any $\gamma^c$ and 
 $\gamma^d$, even if  $c=d$.}.\\
{\bf b.} Relations of Eq.~(\ref{sabtildesab}) 
remain unchanged~\footnote{One easily checks
that $\tilde{\gamma}^{a \dagger} \gamma^c= -i \gamma^c \gamma^{a \dagger}=
-i \eta^{aa}\gamma^c \gamma^a$ $= \eta^{aa} \tilde{\gamma}^a \gamma^c =$ 
$-i \eta^{aa}\gamma^c \gamma^a$.}.\\
{\bf c.} The eigenvalues of the operators $S^{ab}$  and $\tilde{S}^{ab}$ 
on nilpotents and projectors of $\gamma^a$'s are after the reduction of Clifford space
equal to
\begin{eqnarray}
\label{signature0}
S^{ab} \,\stackrel{ab}{(k)} = \frac{k}{2}  \,\stackrel{ab}{(k)}\,,\quad && \quad
\tilde{S}^{ab}\,\stackrel{ab}{(k)} = \frac{k}{2}  \,\stackrel{ab}{(k)}\,,\nonumber\\
S^{ab}\,\stackrel{ab}{[k]} =  \frac{k}{2}  \,\stackrel{ab}{[k]}\,,\quad && \quad 
\tilde{S}^{ab} \,\stackrel{ab}{[k]} = - \frac{k}{2}  \,\,\stackrel{ab}{[k]}\,,
\end{eqnarray}
demonstrating that the eigenvalues of $S^{ab}$ on nilpotents and projectors of 
$\gamma^a$'s differ from the eigenvalues of $\tilde{S}^{ab}$, so that 
$\tilde{S}^{ab}$ can be used to denote irreducible representations of $S^{ab}$
 with the ''family" quantum number, what solves the problems vii. b. and vii. c. 
of Subsect.~\ref{propertiesCliffodd}.\\
{\bf d.} We further recognize 
that $\gamma^a$ transform  $\stackrel{ab}{(k)}$ into  $\stackrel{ab}{[-k]}$, never 
to $\stackrel{ab}{[k]}$, while $\tilde{\gamma}^a$ transform  $\stackrel{ab}{(k)}$ 
into $\stackrel{ab}{[k]}$, never to $\stackrel{ab}{[-k]}$ 
\begin{eqnarray}
&&\gamma^a \stackrel{ab}{(k)}= \eta^{aa}\stackrel{ab}{[-k]},\; \quad
\gamma^b \stackrel{ab}{(k)}= -ik \stackrel{ab}{[-k]}, \; \nonumber\\
&&\gamma^a \stackrel{ab}{[k]}= \stackrel{ab}{(-k)},\;\quad 
\gamma^b \stackrel{ab}{[k]}= -ik \eta^{aa} \stackrel{ab}{(-k)}\,,\nonumber\\
&&\tilde{\gamma^a} \stackrel{ab}{(k)} = - i\eta^{aa}\stackrel{ab}{[k]},\;\quad
\tilde{\gamma^b} \stackrel{ab}{(k)} =  - k \stackrel{ab}{[k]}, \;\nonumber\\
&&\tilde{\gamma^a} \stackrel{ab}{[k]} =  \;\;i\stackrel{ab}{(k)},\; \quad
\tilde{\gamma^b} \stackrel{ab}{[k]} =  -k \eta^{aa} \stackrel{ab}{(k)}\,. 
\label{snmb:gammatildegamma}
\end{eqnarray}
{\bf e.}
One finds, using  Eq.~(\ref{tildegammareduced}),
\begin{eqnarray}
\stackrel{ab}{\tilde{(k)}} \, \stackrel{ab}{(k)}& =& 0\,, 
\quad \;
\stackrel{ab}{\tilde{(-k)}} \, \stackrel{ab}{(k)} = -i \,\eta^{aa}\,  
\stackrel{ab}{[k]}\,,\nonumber\\
\stackrel{ab}{\tilde{(k)}} \, \stackrel{ab}{[k]} &=& i\, \stackrel{ab}{(k)}\,,
\quad\;
\stackrel{ab}{\tilde{(k)}}\, \stackrel{ab}{[-k]} = 0\,, \nonumber\\
%
%
\stackrel{ab}{\tilde{[k]}} \, \stackrel{ab}{(k)}& =& \, \stackrel{ab}{(k)}\,, 
\quad \; 
\stackrel{ab}{\tilde{[-k]}} \, \stackrel{ab}{(k)} = \, 0 \,, \nonumber\\
\stackrel{ab}{\tilde{[k]}} \, \stackrel{ab}{[k]} &=&  0\,,
\quad\;
\stackrel{ab}{\tilde{[- k]}} \, \stackrel{ab}{[k]} =  \, \stackrel{ab}{[k]}\,.
\label{graphbinomsfamilies}
\end{eqnarray}
{\bf f.}
From Eq.~(\ref{snmb:gammatildegamma}) it follows
\begin{eqnarray}
\label{stildestrans}
S^{ac}\stackrel{ab}{(k)}\stackrel{cd}{(k)} &=& -\frac{i}{2} \eta^{aa} \eta^{cc} 
\stackrel{ab}{[-k]}\stackrel{cd}{[-k]}\,,\,\nonumber\\
\tilde{S}^{ac}\stackrel{ab}{(k)}\stackrel{cd}{(k)} &=& \frac{i}{2} \eta^{aa} \eta^{cc} 
\stackrel{ab}{[k]}\stackrel{cd}{[k]}\,,\,\nonumber\\
S^{ac}\stackrel{ab}{[k]}\stackrel{cd}{[k]} &=& \frac{i}{2}  
\stackrel{ab}{(-k)}\stackrel{cd}{(-k)}\,,\,\nonumber\\
\tilde{S}^{ac}\stackrel{ab}{[k]}\stackrel{cd}{[k]} &=& -\frac{i}{2}  
\stackrel{ab}{(k)}\stackrel{cd}{(k)}\,,\,\nonumber\\
S^{ac}\stackrel{ab}{(k)}\stackrel{cd}{[k]}  &=& -\frac{i}{2} \eta^{aa}  
\stackrel{ab}{[-k]}\stackrel{cd}{(-k)}\,,\,\nonumber\\
\tilde{S}^{ac}\stackrel{ab}{(k)}\stackrel{cd}{[k]} &=& -\frac{i}{2} \eta^{aa}  
\stackrel{ab}{[k]}\stackrel{cd}{(k)}\,,\,\nonumber\\
S^{ac}\stackrel{ab}{[k]}\stackrel{cd}{(k)} &=& \frac{i}{2} \eta^{cc}  
\stackrel{ab}{(-k)}\stackrel{cd}{[-k]}\,,\,\nonumber\\
\tilde{S}^{ac}\stackrel{ab}{[k]}\stackrel{cd}{(k)} &=& \frac{i}{2} \eta^{cc}  
\stackrel{ab}{(k)}\stackrel{cd}{[k]}\,. 
\end{eqnarray}
{\bf g.}
Each irreducible representation 
 has now the "family" quantum
number, determined by $\tilde{S}^{ab}$ of the Cartan subalgebra of 
Eq.~(\ref{cartancliff}).
Correspondingly the creation and annihilation operators fulfill algebraically the anticommutation relations of Dirac second quantized fermions: Different irreducible representations carry different 
"family" quantum numbers and to each "family" quantum member only one Hermitian 
conjugated partner with the same "family" quantum number belong. 
Also each summand of the vacuum state,  Eq.~(\ref{vac1}), belongs to a particular
 "family". 
This solves the problem vii. a. of Subsect.~\ref{propertiesCliffodd}. 

The anticommutation relations of Dirac fermions are therefore fulfilled 
on the vacuum state,  Eq.~(\ref{vac1}), on the algebraic level, without 
postulating them.
They follow by themselves from the fact that the creation and 
annihilation operators are superposition of odd products of $\gamma^{a}$'s. 

{\bf Statement 1: } The oddness of the products of  $\gamma^a$'s guarantees the 
anticommuting properties of all objects which include odd number of $\gamma^a$'s.

We shall show in Subsect.~\ref{action} of this section, and in Sect.~\ref{HilbertCliff0},  that the same 
relations are valid also on the Hilbert space of all the second quantized fermions 
states, with the creation operators defined on the tensor product of ''basis vectors''
of the Clifford algebra and on the basis of the momentum space, where the Hilbert 
space is defined with the creation operators of all possible momenta of all possible 
"Slater determinants" applying on $|\psi_{oc}>$. 
%

Let us write down the anticommutation relations of Clifford odd "basic vectors",
representing the creation operators and of the 
corresponding annihilation operators again.
\begin{eqnarray}
\{ \hat{b}^{m}_{f}, \hat{b}^{m' \dagger}_{f'} \}_{*_{A}+}\, |\psi_{oc}> 
&=& \delta^{m m'} \, \delta_{ff'} \,  |\psi_{oc}>\,,\nonumber\\
\{ \hat{b}^{m}_{f}, \hat{b}^{m'}_{f'} \}_{*_{A}+}  \,  |\psi_{oc}>
&=& 0 \,\cdot\,  |\psi_{oc}>\,,\nonumber\\
\{\hat{b}^{m  \dagger}_{f},\hat{b}^{m' \dagger}_{f'}\}_{*_{A}+} \, |\psi_{oc}>
&=& 0 \, \cdot\, |\psi_{oc}>\,,\nonumber\\
 \hat{b}^{m \dagger}_{f} \,{}_{*_{A}} |\psi_{oc}>&=& |\psi^{m}_{f}>\,, \nonumber\\
 \hat{b}^{m}_{f}   \,{*_{A}}  |\psi_{oc}>&=& 0 \,\cdot\,  |\psi_{oc}>\,,
\label{alphagammatildeprod}
\end{eqnarray}
with ($m,m'$) denoting the "family" members and ($f,f'$) denoting "families",
${*_{A}}$ represents the algebraic multiplication of $ \hat{b}^{m}_{f} $ with
the vacuum state $|\psi_{oc}>$ of Eq.(\ref{vac1}) and among themselves, taking 
into account Eq.~(\ref{gammatildeantiher}).\\
{\bf h.} The vacuum state for the vector space determined by $\gamma^a$'s 
remains unchanged $|\psi_{oc}>$, Eq.~(80) of Ref.~\cite{nh2018}, it is a sum of the 
products of any annihilation operator with its Hermitian conjugated partner of any family.
\begin{eqnarray}
|\psi_{oc}>&=& \stackrel{03}{[-i]} \stackrel{12}{[-]} \stackrel{56}{[-]}\cdots
\stackrel{d-1\;d}{[-]} + \stackrel{03}{[+i]} \stackrel{12}{[+]} \stackrel{56}{[-]} \cdots
               \stackrel{d-1\;d}{[-]} \nonumber\\
  + &&\stackrel{03}{[+i]} \stackrel{12}{[-]} \stackrel{56}{[+]}\cdots
\stackrel{d-1\;d}{[-]} + \cdots |1>\,, \quad \nonumber\\
&&{\rm for}\; d=2(2n+1)\,,
\nonumber\\
|\psi_{oc}>&=& \stackrel{03}{[-i]} \stackrel{12}{[-]} \stackrel{35}{[-]}\cdots
               \stackrel{d-3\;d-2}{[-]}\stackrel{d-1\;d}{[+]} \nonumber\\
  + &&\stackrel{03}{[+i]} \stackrel{12}{[+]} 
\stackrel{56}{[-]}
\cdots \stackrel{d-3\;d-2}{[-]} \;\,\stackrel{d-1\;d}{[+]} + \cdots |1>\,, \quad  \nonumber\\
&&{\rm for}\; d=4n\,,
\label{vac1}
\end{eqnarray}
$n$ is a positive integer.\\
%
%
%
{\bf i.}
Taking into account the relations among $\theta^a$, 
$\frac{\partial}{\partial \theta_{a}}$, $\gamma^a$ and $\tilde{\gamma}^a$, 
presented in Eq.~(\ref{clifftheta}), to express $\gamma^a$ and $\tilde{\gamma}^a$
with $\theta^a$ and $\frac{\partial}{\partial \theta_{a}}$, as well as 
Eq.~(\ref{tildegammareduced}) requiring that 
\begin{eqnarray}
\label{partialthetazero}
&&[\tilde{\gamma}^a (a_0 + a_{bc}\gamma^b \gamma^c + a_{bcde}\gamma^b
 \gamma^c \gamma^d \gamma^e +\cdots) =  i  (a_0 + a_{bc}\gamma^b \gamma^c + a_{bcde}\gamma^b \gamma^c \gamma^d \gamma^e +\cdots) \gamma^a ]
 |\psi_{oc}>\,, \nonumber\\
&&[\tilde{\gamma}^a (a_{b}\gamma^b + a_{bcd}\gamma^b
\gamma^c \gamma^d +\cdots) = - i  (a_{b}\gamma^b + a_{bcd}\gamma^b
 \gamma^c \gamma^d + \cdots) \gamma^a ]
|\psi_{oc}>\,, \nonumber
\end{eqnarray}
one obtains 
$\frac{\partial}{\partial \theta_{a}}\Rightarrow 0 $ as the only solution.  Then it follows
\begin{eqnarray}
\theta^a \Rightarrow \gamma^a\,, 
\label{tildegammareduced1}
\end{eqnarray}
which does not mean that $\theta^a$ is equal to $\gamma^a$ but rather that the whole 
Grassmann algebra reduces to only one of the two Clifford algebras, the one, in which
the ''basis vectors''  are superposition of products of (odd when describing fermions) 
number of $\gamma^a$'s.

Eq.~(\ref{tildegammareduced})) namely requires: 
  $\tilde{\gamma}^a (a_0 + a_b\gamma^b + 
  a_{bc} \gamma^b \gamma^c+ \cdots ) =
 ( i a_0 \gamma^a + (-i) a_b  \gamma^b \gamma^a + 
  i a_{bc} \gamma^b \gamma^c  \gamma^a +\cdots)$. 
  
  Checking the relations presented in Eq.~(\ref{gammatildeantiher}) one finds that
  all  the relations of this equations remain valid also after the reduction of the 
  Clifford space~\footnote{
 Let us check
  $\{ \tilde{\gamma}^{a}, \tilde{\gamma}^{b}\}_{+}= 2\eta^{ab}=$
$\tilde{\gamma}^{a} \tilde{\gamma}^{b}+\tilde{\gamma}^{b}
\tilde{\gamma}^{a}=$ $ \tilde{\gamma}^{a} i\gamma^b +\tilde{\gamma}^{b} i \gamma^a=$
 $ i \gamma^b (-i)\gamma^a + i\gamma^a(-i)\gamma^b= 2\eta^{ab} $.
 $\{ \tilde{\gamma}^{a}, \gamma^b\}_{+}= 0=$
 $\tilde{\gamma}^{a} \gamma^b+\gamma^b \tilde{\gamma}^{a}=$
 $ \gamma^b (-i)\gamma^a+  \gamma^b i \gamma^a=0$.
For a particular case one has $\{ \tilde{\gamma}^{a}, \gamma^a\}_{+}= 0=$
$ \tilde{\gamma}^{a} \gamma^a + \gamma^a \tilde{\gamma}^a= $
 $\gamma^a (-i )\gamma^a +\gamma^a i \gamma^a=0$.}.
The application of $\tilde{\gamma}^a$  depends on the space on which it applies,
Eq.~(\ref{tildegammareduced}).

The Hermitian conjugated part of the space in the Grassmann case is "freezed out"
together with the "vector" space of $\tilde{\gamma}^a$'s.

\subsection{Clifford fermions with families
}
\label{Clifffamilies}
%


Let us make a choice of  the starting creation operator $\hat{b}^{1 \dagger}_{1}$ 
of an odd Clifford character and of its Hermitian conjugated partner in $d=2(2n+1)$ and $d=4n$, respectively, as follows 
%
\begin{eqnarray}
{\hat b}^{1 \dagger}_{1}\, {\bf :} &=& \stackrel{03}{(+i)} \stackrel{12}{(+)} 
\stackrel{56}{(+)}\cdots 
\stackrel{d-3\;d-2}{(+)}\;\;\stackrel{d-1\;d}{(+)}\,,\nonumber\\
({\hat b}^{1 \dagger}_{1})^{\dagger} &=& {\hat b}^{1}_{1}\, {\bf :}= 
\stackrel{d-1\;d}{(-)} \;\;
\stackrel{d-3\;d-2}{(-)}\cdots  \stackrel{56}{(-)}  \stackrel{12}{(-)}  \stackrel{01}{(-i)}\,,
\nonumber\\
d&=&2(2n+1)\,,
 \nonumber\\
{\hat b}^{1 \dagger}_{1} \, {\bf :} &=& \stackrel{03}{(+i)} \stackrel{12}{(+)} 
\stackrel{56}{(+)}\cdots 
\stackrel{d-3\;d-2}{(+)}\;\;\stackrel{d-1\;d}{[+]}\,,\nonumber\\
 ({\hat b}^{1 \dagger}_{1})^{\dagger}&=& {\hat b}^{1}_{1}\, {\bf :} = 
\stackrel{d-1\;d}{[+]} \;\;
\stackrel{d-3\;d-2}{(-)}\cdots  \stackrel{56}{(-)}  \stackrel{12}{(-)}  \stackrel{01}{(-i)}\,,
\nonumber\\
d&=&4n\,.
\label{start(2n+1)2cliffgammatilde4n}
\end{eqnarray}
 All the rest "vectors", belonging to the same Lorentz representation, follow by
the application of the Lorentz generators $S^{ab}$'s. 


The representations with different "family" quantum numbers are reachable by
$\tilde{S}^{ab}$, since, according to Eq.~(\ref{stildestrans}), we recognize that $\tilde{S}^{ac}$
transforms two nilpotents $ \stackrel{ab}{(k)} \stackrel{cd}{(k)}$ into two projectors 
$ \stackrel{ab}{[k]} \stackrel{cd}{[k]}$, without changing $k$ ($\tilde{S}^{ac}$ 
transforms $ \stackrel{ab}{[k]} \stackrel{cd}{[k]}$ into $ \stackrel{ab}{(k)} \stackrel{cd}{(k)}$, 
as well as $ \stackrel{ab}{[k]} \stackrel{cd}{(k)}$ into $ \stackrel{ab}{(k)} \stackrel{cd}{[k]}$). 
All the "family" members are reachable from one member of a new family by the 
application of $S^{ab}$'s.

In this way, by starting with the creation operator ${\hat b}^{1 \dagger}_{1}$, 
Eq.~(\ref{start(2n+1)2cliffgammatilde4n}), $2^{\frac{d}{2}-1}$ "families", each with 
$2^{\frac{d}{2}-1}$ "family" members follow. 

Let us find the starting member of the next "family" to the "family" of 
Eq.~(\ref{start(2n+1)2cliffgammatilde4n}) by the application of $\tilde{S}^{01}$
\begin{eqnarray}
{\hat b}^{1 \dagger}_{2}\, {\bf :} &=& \stackrel{03}{[+i]} \stackrel{12}{[+]} 
\stackrel{56}{(+)}\cdots \stackrel{d-3\;d-2}{(+)}\;\;\stackrel{d-1\;d}{(+)}\,,\nonumber\\
{\hat b}^{1}_{2}\,{\bf :} &=& 
\stackrel{d-1\;d}{(-)} \;\;
\stackrel{d-3\;d-2}{(-)}\cdots  \stackrel{56}{(-)}  \stackrel{12}{[+]}  \stackrel{01}{[+i]}\,.
\label{d=2(2n+1)}
\end{eqnarray}


The corresponding annihilation operators, that is the Hermitian conjugated partners of 
$2^{\frac{d}{2}-1}$ "families", each with $2^{\frac{d}{2}-1}$ "family" members, 
following from the starting creation operator ${\hat b}^{1 \dagger}_{1}$ by the application of $S^{ab}$'s --- the family members --- and the application of 
 $\tilde{S}^{ab}$ --- the same family member of another family --- 
can be obtained by Hermitian conjugation.

{\it The creation and annihilation operators of an odd Clifford character, expressed by 
nilpotents and projectors of  $\gamma^a$'s, obey  anticommutation relations of}$\,$ 
Eq.~(\ref{alphagammatildeprod}),
{\it without postulating the second quantized anticommutation relations} as we explain 
in Subsect.~\ref{postulates}.

The even partners of the Clifford odd creation and annihilation operators follow by either the 
application of $\gamma^a$ on the creation operators, leading to  $2^{\frac{d}{2}-1}$
 "families", each with $2^{\frac{d}{2}-1}$ members,  or with the application of 
$\tilde{\gamma}^a$ on the creation  operators,  leading to another group of the 
Clifford even operators, again with the  $2^{\frac{d}{2}-1}$ "families", each with 
$2^{\frac{d}{2}-1}$ members.

It is not difficult to recognize, that each of the Clifford even "families", obtained by  
the application of $\gamma^a$ or by $\tilde{\gamma}^a$ on the creation operators, 
contains one selfadjoint operator, which is the product  of projectors only, contributing 
as a summand to the vacuum state, Eq.~(\ref{vac1}).





%
 \subsection{Action for free massless Clifford fermions with half integer spin and solutions of Weyl equations}
%
\label{action}

To relate the creation operators, expressed with the Clifford odd ''basis vectors'',
and the creation operators, creating the second quantized fermions, we define 
the tensor products of the finite number of odd Clifford ''basis vectors'' and infinite 
basis  of momentum space. To compare properties of our creation operators of the 
second quantized fermions with those of Dirac, the solution of the equations of  
motion of the Weyl (for massless free fermions) or of the Dirac  equations are 
appropriate. 

The Lorentz invariant action for a free massless fermion in Clifford space is well known 
\begin{eqnarray}
{\cal A}\,  &=& \int \; d^dx \; \frac{1}{2}\, (\psi^{\dagger}\gamma^0 \, \gamma^a p_{a} \psi) +
 h.c.\,, 
\label{actionWeyl}
\end{eqnarray}
$p_{a} = i\, \frac{\partial}{\partial x^a}$, leading to the equation of motion 
\begin{eqnarray}
\label{Weyl}
\gamma^a p_{a}  |\psi>&= & 0\,, 
\end{eqnarray}
and to the Klein-Gordon equation
\begin{eqnarray}
\label{LtoKG}
\gamma^a p_{a} \gamma^b p_b |\psi>&= &   
p^a p_a |\psi>=0\,,\nonumber
\end{eqnarray}
%
$\gamma^0$ appears in the 
action to take care of the Lorentz invariance of the action.

Our Clifford algebra ''basis vectors'' offer the description of only the internal degrees of freedom of fermions (in $d=(3+1)$ the ''basis vectors'' offers the description of only the spin and family degrees of freedom, in $d\ge 5$ also of the charges~\cite{gn2009,IARD2016,n2014matterantimatter,normaJMP2015} and the references therein).

We need to extend the internal degrees of freedom 
(offering final number ---  $2^{\frac{d}{2}-1} \times 2^{\frac{d}{2}-1}$ --- of basis 
vectors of the odd products of $\gamma^a$) to the momentum or coordinate space 
with (infinite number of) basis.  

{\bf Statement 2:} For deriving the anticommutation relations for the Clifford fermions,
to be compared with the anticommutation relations of the second quantized fermions, 
we need to define the tensor product of the Clifford odd ''basis vectors''  and the 
momentum space
\begin{eqnarray}
{\bf basis}_{(p^a, \gamma^a )} = |p^a>\,*_{T}\, |\gamma^a>\,.                                                                                                    
 \nonumber
 \end{eqnarray}

The new state vector space is the tensor product of the internal space of fermions
and the space of momenta or coordinates. All states have an odd Clifford character
due to oddness of the internal space.  

Solutions  of Eq.~(\ref{Weyl}) for free massless 
fermions of momentum $p^a, a=(0,1,2,3,5,\dots,d)$ are superposition of 
''basis vectors'' ${\hat b}^{m \dagger}_{f}$, expressed by operators $\gamma^a$,
where $f$ denotes a "family" and $m$ a "family" member quantum number, 
Eqs.~(\ref{start(2n+1)2cliffgammatilde4n}, \ref{d=2(2n+1)}),
and of plane waves in the case of free, in our case, massless fermions. The 
equations of motion require that $|p^0|=|\vec{p}|$. Correspondingly it follows 
%
\begin{eqnarray}
<x|\bf {\psi^{s f}} (\vec{p}, p^0)>|_{p^0=|\vec{p}|}  
&=& \int dp^0 \delta(p^0 -|\vec{p}|) \,\hat{\bf b}^{s f \dagger}(\vec{p}) \, 
e^{-i p_a x^a} \, {}*_{A}|\psi_{oc}>\nonumber\\
&=&( {\hat{\bf b}}^{ s f \dagger} (\vec{p})
\cdot \,e^{-i (p^0 x^0 - \varepsilon \vec{p}\cdot \vec{x})})|_{p^0=|\vec{p}|}
 \,{}*_{A}\,|\psi_{oc}>\,,
\nonumber\\
%
&& {\rm where\;we \;define}
\,,\nonumber\\
{\hat{\bf b}}^{ s f \dagger} (\vec{p})|_{p^0=|\vec{p}|}\, 
& \stackrel{\mathrm{def}}{=}& 
\sum_{m} c^{ s f}{}_{m}\; (\vec{p},|p^0|=|\vec{p}|) \, \,
\hat{b}^{ m \dagger}_{f}\,, 
\nonumber\\ 
|\bf{\psi^{s f}} (\vec{x},x^0)>&=&  \int_{- \infty}^{+ \infty} \,
\frac{d^{d-1}p}{(\sqrt{2 \pi})^{d-1}} \, (\hat{\bf b}^{s f \dagger}(\vec{p})\, 
e^{-i (p^0 x^0- \varepsilon \vec{p}\cdot \vec{x})}|_{p^0=|\vec{p}|} \,{}*_{A}\,|\psi_{oc}>\,,
\label{Weylp}
\end{eqnarray}
$s$ represents different orthonormalized solutions of the equations of motion, 
$\varepsilon=\pm1$, depending on handedness and spin of solutions, 
$c^{ s f}{}_{m} (\vec{p},|p^0|=|\vec{p}| )$ are coefficients, depending on 
momentum $|\vec{p}|$ with $|p^0|=|\vec{p}|$, while ${}*_{A}$ denotes 
the algebraic multiplication of the ''basis vectors''
$\hat{b}^{ m \dagger}_{f}$ on the vacuum state $|\psi_{oc}>$, Eq.~(\ref{alphagammatildeprod}).

An illustration of 
${\hat{\bf b}}^{ s f \dagger} (\vec{p})$ is presented in Subsect.~\ref{illustration}. 

Since the ``basis vectors'' in internal space of fermions are orthogonal according to 
Eq.~(\ref{almostDirac}) ($\{\hat{b}^{ m}_{f}\,{}_{*_{A}}\,,\,
 \hat{b}^{ m' \dagger}_{f'}\,{}_{*_{A}}\}_{+}|\psi_{oc}>=$
 $\hat{b}^{ m}_{f}\;{}_{*_{A}}\, \hat{b}^{ m' \dagger}_{f'}\,{}_{*_{A}}|\psi_{oc}>$), 
\begin{eqnarray}
&&\hat{b}^{ m}_{f}\;{}_{*_{A}}\, \hat{b}^{ m' \dagger}_{f'}\,{}_{*_{A}}|\psi_{oc}> = \delta^{m m'} \delta_{f f'}\,|\psi_{oc}>\,,
\nonumber\\
&&{\rm it \; follows \;for\; particular\;}  \vec{p}\,, p^0=|\vec{p}|,{\;\rm that} \nonumber\\
&& \sum_{m} c^{s f *}{}_{m}  (\vec{p},|p^0|=|\vec{p}| )\;\; 
c^{s' f'}{}_{m} (\vec{p},|p^0|=|\vec{p}| ) =
 \delta^{s s'} \delta_{f f'}\, \,,\nonumber\\
&& {\rm leading \; to} \nonumber\\
&&\int\,  \frac{d^{d-1}x}{(\sqrt{2\pi})^{d-1}}  \,<{\bf \psi^{s' f'}} ( \vec{p'}, p'^0=|\vec{p'}|)|x>\,
<x| |\,{\bf \psi^{s f}}(\vec{p}, p^0
=|\vec{p}|)>=\,\nonumber\\
&& \int \frac{d^{d-1}x}{(\sqrt{2\pi})^{d-1}} 
e^{i p'_a x^a}|_{p'^0=|\vec{p'}|} \, 
e^{-i p_a x^a}|_{p^0|=|\vec{p}|} \, \nonumber\\
&&\cdot \,<\psi_{oc}| \,(\hat{\bf b}^{s' f'} (\vec{p'}) \,
 \,\hat{\bf b}^{s f \dagger} (\vec{p}))\,{}_{*_{A}} \, 
 |\psi_{oc}>=\,
\delta^{}{s s'} \delta^{f f'} \delta(\vec{p}- \vec{p'})\,,
%
\label{Weylpp'solort}
\end{eqnarray}
%
while we 
take into account
that $\int \frac{d^{d-1}x}{(\sqrt{2\pi})^{d-1}}  \, e^{i p'_a x^a} \, 
 e^{-i p_a x^a} = \delta(\vec{p}-\vec{p'})$.
 
 Let us now evaluate the scalar product $<{\bf \psi^{s f}} (\vec{x}, x^0)\,
  |\,{\bf \psi^{s' f'}} (\vec{x}{\,}', x^0)>$, taking into account that the scalar 
  product is evaluated at a time  $x^0$ and correspondingly using the relation
 \begin{eqnarray}
&&<{\bf \psi^{s f}} ( \vec{x}, x^0) |\,{\bf \psi^{s' f'}}(\vec{x}{\,}', x^0)>=
\delta^{s s'}\, \delta_{f f'} \,\delta( \vec{x} - \vec{x}{\,}') =\nonumber\\
&& \int \frac{dp^0}{\sqrt{2\pi}}\, \int \frac{dp'^0}{\sqrt{2\pi}} 
\delta(p^0 -p'^0) \,\int_{- \infty}^{+ \infty} \,
\frac{d^{d-1}p'}{(\sqrt{2 \pi})^{d-1}} \, 
\,\int_{- \infty}^{+ \infty} \,
\frac{d^{d-1}p}{(\sqrt{2 \pi})^{d-1}} \delta(p^0 -|\vec{p}|)\,
\delta(p'^0 -|\vec{p'}|)\,\nonumber\\ 
&&<\psi_{oc}| ( \hat{\bf {b}}^{s' f'}(\vec{p'}, p'^0) \,
  \hat{\bf b}^{s f \dagger}(\vec{p}, p^0) )\,{}_{*_{A}}\, |\psi_{oc}>
 e^{i p'_a x'^a}\, e^{-i p_a x^a} =\nonumber\\
 &&\int \frac{dp^0}{\sqrt{2\pi}}\,
\int_{- \infty}^{+ \infty} \frac{d^{d-1}p'}{(\sqrt{2 \pi})^{d-1}} \, 
\delta(p^0 -|\vec{p'}|)
\int_{- \infty}^{+ \infty} \frac{d^{d-1}p}{(\sqrt{2 \pi})^{d-1}} \, 
\delta(p^0 -|\vec{p}|)\,\nonumber\\
&&<\psi_{oc}| (\hat{\bf {b}}^{s f}(\vec{p}, p^0) \,
  \hat{\bf b}^{s' f'\dagger}(\vec{p'}, p^0))\,{}_{*_{A}} \, |\psi_{oc}>
 e^{i (p^0x^0-\vec{p}\cdot \vec{x})}\,  
 e^{-i (p^0x'^0-\vec{p'}\cdot \vec{x})}\,.
\label{xx'scalarprod}
\end{eqnarray}
The scalar product 
$<{\bf \psi^{s f}} (\vec{x}, x^0)\,  |\,{\bf \psi^{s' f'}} (\vec{x}{\,}', x^0)>$ has 
obviously the desired properties of the second quantized states.

\vspace{1mm}

Let us define the creation operators $ \underline{\hat{\bf b}}^{s f  \dagger}_{tot}(\vec{p})$, which determine, when applying on the vacuum state, 
the fermion states, Eq.~(\ref{Weylp}), 
\begin{eqnarray}
\label{creationb}
\underline{\hat{\bf b}}^{s f \dagger}_{tot}(\vec{p}) & \stackrel{\mathrm{def}}{=}&
{\hat {\bf b}}^{s f \dagger}(\vec{p})\, e^{-i (p^0 x^0 - \vec{p}\cdot \vec{x})}\,,\nonumber\\
\underline{\hat{\bf b}}^{s f}_{tot}(\vec{p}) &=&
(\underline{\hat{\bf b}}^{s f \dagger}_{tot}(\vec{p}))^{\dagger} =
{\hat {\bf b}}^{s f}(\vec{p})\, e^{i (p^0 x^0 - \vec{p}\cdot \vec{x})}\,,
\nonumber\\
\underline{\hat{\bf b}}^{s f \dagger}_{tot}(\vec{p})\, |\psi_{oc}>&=&
|\,{\bf \psi^{s f}}(\vec{p}, p^0=|\vec{p}|)>\,. 
\end{eqnarray}
%
In Eq.~(\ref{creationb}) 
$\underline{\hat{\bf b}}^{s f \dagger}_{tot}(\vec{p})$  creates 
on the vacuum state $|\psi_{oc}>$ the single fermion states. 
We can multiply, using  the tensor product $*_{T}$ multiplication this time,
an arbitrary number of such single particle states, what means 
that we multiply an arbitrary number of creation operators 
$\underline{\hat{\bf b}}^{s f \dagger}_{tot}(\vec{p})*_{T}$
$\underline{\hat{\bf b}}^{s' f' \dagger}_{tot}(\vec{p'})*_{T}\cdots$
$*_{T}\underline{\hat{\bf b}}^{s'' f'' \dagger}_{tot}(\vec{p''})$, 
applying on $|\psi_{oc}>$, which gives nonzero contributions, provided that they distinguish among themselves
in at least one of the properties $(s,f,\vec{p})$, in the internal space quantum 
numbers $(s,f)$ or in momentum part $\vec{p}$, due to the orthonormal property
of plane waves.

The space of all such functions, which one can form - including the identity - 
represents the second quantized Hilbert space.
We present these tensor products as ''Slater determinants'' of occupied and empty
states  in Section~\ref{HilbertCliff0}.

Due to  anticommutation relations of any two of creation operators  
$\{{\hat {\bf b}}^{s f\dagger}(\vec{p})\,,\,{\hat {\bf b}}^{s' f'}(\vec{p})\}_{+}
\, |\psi_{oc}> =\delta^{f f'} \delta^{s s'} \, |\psi_{oc}>$, Eqs.~(\ref{alphagammatildeprod}, \ref{Weylp}),
while plane waves form the orthonormal basis in the momentum representation, Eq.~(\ref{Weylpp'solort}), the new creation operators 
$\underline{\hat{\bf b}}^{s f \dagger}_{tot}(\vec{p})$, which are are generated 
on the tensor products of both spaces, internal and momentum, fulfill the anticommutation 
relations when applied on $|\psi_{oc}>$. 
\begin{eqnarray}
\{{\underline {\hat{\bf b}}}_{tot}^{s f} (\vec{p})\, , {\underline {\hat{\bf b}}}_{tot}^{s f \dagger} (\vec{p}{\,}') \}_{+} \,{}_{*_{T}}|\psi_{oc}>\,
&=& \delta^{s s'} \, \delta_{f f'} \, \delta(\vec{p} - \vec{p'})\,  |\psi_{oc}>\,,\nonumber\\
\{{\underline {\hat{\bf b}}}_{tot}^{s f} (\vec{p})\,, {\underline {\hat{\bf b}}}_{tot}^{s' f'} (\vec{p'}) \}_{+} \,{}_{*_{T}}|\psi_{oc}>\,
&=& 0 \, \cdot |\psi_{oc}>\,,\nonumber\\
\{{\underline {\hat{\bf b}}}_{tot}^{s f \dagger} (\vec{p}) \,,{\underline {\hat{\bf b}}}_{tot}^{s' f' \dagger} (\vec{p}{\,}') \}_{+} \,{}_{*_{T}} |\psi_{oc}>\,
&=& 0 \, \cdot |\psi_{oc}>\,,\nonumber\\
{\underline {\hat{\bf b}}}_{tot}^{s f \dagger} (\vec{p})  \,{}_{*_{T}} |\psi_{oc>} 
&=& |{\psi^{s f}} ( \vec{p})>\,, \nonumber\\
 {\underline {\hat{\bf b}}}_{tot}^{s f } (\vec{p})  \,{}_{*_{T}} |\psi_{oc}>
&=& 0 \,\cdot |\psi_{oc}>\,, \nonumber\\
 |p^0|& =& |\vec{p}|\,.
\label{Weylpp'comrel}
\end{eqnarray}
It is not difficult to show that  ${\underline {\hat{\bf b}}}_{tot}^{s f} (\vec{p})$
and ${\underline {\hat{\bf b}}}_{tot}^{s f \dagger} (\vec{p})$  manifest the same
anticommutation relations also on tensor products of an arbitrary chosen set of
single fermion states, what we discuss in Sect.~\ref{HilbertCliff0}.

Therefore, with the choice of the Clifford odd ''basis states'' to describe the internal 
space of fermions (we can proceed equivalently in the Grassmann case)  and using 
the tensor product  of the internal space and the momentum or coordinate space 
to solve the equations of motion, we derive the anticommutation relations among 
creation operators 
$ \underline{\hat{\bf b}}^{s f \dagger}_{tot}(\vec{p})$ and their Hermitian conjugated partners annihilation operators $ (\underline{\hat{\bf b}}^{s f \dagger}_{tot}(\vec{p}))^{\dagger}=$ ${\hat {\bf b}}^{s f }(\vec{p}) \,e^{i (p^0x^0-\vec{p,}\cdot \vec{x})} =$ $ \underline{\hat{\bf b}}^{s f}_{tot}(\vec{p})$, with $|p^0| = |\vec{p}|$.  While application of $ \underline{\hat{\bf b}}^{s f \dagger}_{tot}(\vec{p})$ on 
$|\psi_{oc}>$ generates the single fermion state, the application of 
$ \underline{\hat{\bf b}}^{s f}_{tot}(\vec{p})$ gives zero.

We shall demonstrate in Sect.~\ref{HilbertCliff0} that there is 
$\{\underline{\hat{\bf b}}^{s' f'}_{tot}(\vec{p'})
\, , \, \underline{\hat{\bf b}}^{s f \dagger}_{tot}(\vec{p})\}_{+}$, which when 
applied on the Hilbert space of the second quantized fermions (that is on tensor 
products of all single fermion states, or equivalently on all possible ''Slater 
determinants''),  gives zero when at least one of 
$(s',f',\vec{p'})$ differ from $(s,f,\vec{p})$, while 
$\{\underline{\hat{\bf b}}^{s f}_{tot}(\vec{p})
\, , \, \underline{\hat{\bf b}}^{s f \dagger}_{tot}(\vec{p})\}_{+}$ applied 
on the Hilbert space, gives the whole Hilbert space back.

Taking into account the last line of Eq.~(\ref{Weylp}) and Eqs.~(\ref{xx'scalarprod},\ref{creationb}), 
the creation operators ${\underline {\bf {\Huge \Psi}}}^{\dagger}$ follow,
which determine, when applying on the vacuum state $|\psi_{oc}>$, the fermion
fields $|\bf{\psi^{s f}} (\vec{x},x^0)>$, depending on coordinates at particular 
time  $x^0$
\begin{eqnarray}
\label{creationpsi}
{\underline {\bf {\Huge \Psi}}}^{s f \dagger}(\vec{x}, x^0) 
& \stackrel{\mathrm{def}}{=}&  
\int_{- \infty}^{+ \infty} \, \frac{d^{d-1}p}{(\sqrt{2 \pi})^{d-1}} \,
\underline{\hat{\bf b}}^{s f \dagger}_{tot}(\vec{p})_{ |p^0|= |\vec{p}|} \,,
\nonumber\\
{\underline {\bf {\Huge \Psi}}}^{s f \dagger}(\vec{x}, x^0)\, |\psi_{oc}>&=&
 |\,{\bf \psi^{s f}} (\vec{x}, x^0)>\,,\nonumber\\
\{ {\underline {\bf {\Huge \Psi}}}^{s f \dagger}(\vec{x}, x^0)\,, 
{\underline {\bf {\Huge \Psi}}}^{s' f'}(\vec{x'}, x^0) \}_{+}\,  |\psi_{oc}>&=& 
\delta^{s s'} \delta^{f f'} \delta(\vec{x} - \vec{x'})  |\psi_{oc}>\,, \nonumber\\
\{ {\underline {\bf {\Huge \Psi}}}^{s f}(\vec{x}, x^0)\,, 
{\underline {\bf {\Huge \Psi}}}^{s' f'}(\vec{x'}, x^0) \}_{+} \, |\psi_{oc}>&=& 
0\,.\nonumber\\
\{ {\underline {\bf {\Huge \Psi}}}^{s f \dagger}(\vec{x}, x^0)\,, 
{\underline {\bf {\Huge \Psi}}}^{s' f' \dagger}(\vec{x'}, x^0) \}_{+}\,  |\psi_{oc}>&=& 0\,,
\end{eqnarray}
where ${\underline {\bf {\Huge \Psi}}}^{\dagger}(\vec{x}, x^0)$ and 
${\underline {\bf {\Huge \Psi}}}^{s f} (\vec{x}, x^0)$ are creation and annihilation partners, respectively,  Hermitian conjugated to each other, in the coordinate representation, presenting the creation 
and annihilation operators of the second quantized fields.

The application of the creation operators  $\underline{\hat{\bf b}}^{s f \dagger}_{tot}(\vec{p})_{ |p^0|= |\vec{p}|}$ and ${\underline {\bf {\Huge \Psi}}}^{\dagger}(\vec{x}, x^0)$ and their Hermitian conjugated partners  on the Hilbert space of fermion fields will be discussed in Sect.~\ref{HilbertCliff0}.

Dirac uses the  Lagrange and Hamilton formalism for fermion fields  and assuming that the 
second quantized states should anticommute to describe fermions, he derives the anticommuting creation and annihillation operators. 
 In Subsect.~\ref{bethenormarelation} we compare the Dirac anticommutation relations 
 with our way of deriving anticommutation relations for second quantized fields  in details. 


%


%
%

In Subsect.~\ref{illustration} the properties of creation and annihilation operators, 
${\underline {\hat {\bf b}}^{s f\dagger}_{tot}}(\vec{p})$ and 
${\underline {\hat {\bf b}}^{s f}_{tot}}(\vec{p})$, respectively, described by 
the odd Clifford algebra objects in $d=(5+1)$-dimensional 
space are discussed.

%

\subsection{ Illustration of Clifford fermions with families in $d=(5+1)$-dimensional space}
\label{illustration}
\begin{small}
We illustrate properties of the Clifford odd, and correspondingly anticommuting,  
creation and their Hermitian conjugated partners annihilation operators, belonging 
to $2^{\frac{6}{2}-1}=4$ "families", each with  $2^{\frac{6}{2}-1}=4$ members in 
$d=(5+1)$-dimensional space. The spin in  the fifth and the sixth dimension manifests
as the charge in $d=(3+1)$.
%


In Table~\ref{Table Clifffourplet.} the "basis vectors" of  odd and even Clifford character 
 are presented. 
They are "eigenvectors" of the Cartan subalgebras, Eq.~(\ref{cartancliff}). 

Half of the Clifford odd "basis vectors" are (chosen to be) creation operators 
${\hat b}^{m \dagger}_{f}$, denoted 
in table by {\it odd I}, appearing in four "families", $f=(1(a),2(b),3(c),4(d))$. 
The rest half of the Clifford odd "basis vectors" are their Hermitian conjugated partners
 ${\hat b}^{m}_f$, presented in {\it odd II} part and denoted with the 
corresponding "family" and family members ($a_m,b_m,c_m,d_m$) quantum numbers.

The normalized vacuum state is the product of ${\hat b}^{m}_f \cdot$ 
${\hat b}^{m \dagger}_{f}$ --- this product is the same for each member of a 
particular family and different for different families --- summed over four families
 \begin{eqnarray}
\label{vac5+1}
 |\psi_{oc}> &=&\frac{1}{\sqrt{2^{\frac{6}{2}-1}}}\,
 (\stackrel{03}{[-i]}\stackrel{12}{[-1]}\stackrel{56}{[-1]}\
+ \stackrel{03}{[+i]}\stackrel{12}{[+1]}\stackrel{56}{[-1]}
                 \nonumber\\
&+&\stackrel{03}{[+i]}\stackrel{12}{[-1]}\stackrel{56}{[+1]}
+ \stackrel{03}{[-i]}\stackrel{12}{[+1]}\stackrel{56}{[+1]}).
\end{eqnarray}

One easily checks, by taking into account Eq.~(\ref{graphbinomsfamilies}), that  
the creation operators ${\hat b}^{m \dagger}_{f}$
and the annihilation operators ${\hat b}^{m}_{f}$ fulfill the 
anticommutation relations of Eq~(\ref{alphagammatildeprod}).

The summands of the vacuum state $ |\psi_{oc}>$ appear among selfadjoint
members of {\it even I}  part of Table~\ref{Table Clifffourplet.}, each of summands 
belong to different "family"~\footnote{If we would make a choice for creation operators
the  "families" with the ``family'' members of {\it odd II} of Table~\ref{Table Clifffourplet.}, instead of "families" with the ``family'' members of 
{\it odd I}, then their Hermitian conjugated partners would be the "families" with the ``family'' members  in {\it odd I}.
The vacuum state would be  the sum of  products of  annihilation operators of {\it odd I} 
times the creation operators of {\it odd II} and would be the sum of   selfadjoint members appearing in {\it even II}.}.

All the Clifford even "families" with "family" members of  Table~\ref{Table Clifffourplet.} 
can be obtained as algebraic products, $*_{A}$, of the Clifford odd "vectors" of the same table. 


\begin{table*}
\begin{tiny}
\caption{\label{Table Clifffourplet.}  $2^d=64$ "eigenvectors" of the Cartan subalgebra, 
Eq.~(\ref{cartancliff}), of the Clifford  odd and even algebras in $d=(5+1)$ are 
presented, divided into four groups, each group with four "families", each ''family'' with
four ''family'' members. Two of four groups are  sums of an odd number of $\gamma^a$'s.  The "basis vectors", ${\hat b}^{m \dagger}_f$, 
Eqs.~(\ref{start(2n+1)2cliffgammatilde4n}, \ref{d=2(2n+1)}), in  {\it odd I} group, 
belong to four 
"families" ($f=1 (a),2(b),3(c),4(d)$) with four members ($m=1,2,3,4$), having their 
Hermitian conjugated partners, ${\hat b}^{m}_f$, among "basis vectors" of the
{\it odd II} part, denoted by the 
corresponding "family" and "family" members ($a_m,b_m,c_m,d_m$) quantum numbers.
The "family" quantum  numbers, the eigenvalue of $(\tilde{S}^{03}, \tilde{S}^{12},\tilde{S}^{56})$, of ${\hat b}^{m \dagger}_f$ are written   above each "family".
The  two groups with the even number of $\gamma^a$'s, {\it even I} and {\it even II}, 
have their Hermitian conjugated partners within their own group each. There are members in each group, which are products of projectors only. 
Numbers --- 
$\stackrel{03}{\;\,}\;\;\,\stackrel{12}{\;\,}\;\;\,\stackrel{56}{\;\,}$ ---
denote the indexes of the corresponding Cartan subalgebra members 
($\tilde{S}^{03}, \tilde{S}^{12}, \tilde{S}^{56}$), Eq.~(\ref{cartancliff}). 
In the columns $(7,8,9)$ the eigenvalues of the Cartan subalgebra members 
$(S^{03}, S^{12}, S^{56})$, Eq.~(\ref{cartancliff}), are presented. The last two 
columns  tell the handedness of $d=(5+1)$, $\Gamma^{(5+1)}$, and of $d=(3+1)$,
 $\Gamma^{(3+1)}$, respectively, defined in Eq.(\ref{hand}).
}
  \begin{tabular}{|c|c|c|c|c|c|r|r|r|r|r|}
\hline
$ odd \, I$&$m$&$ f=1 (a)$&$ f=2 (b)$&$ f=3 (c)$&
$ f=4 (d)$&$S^{03}$&$S^{12}$&$S^{56}$&$\Gamma^{(5+1)}$&
$\Gamma^{(3+1)}$\\
&&$(\frac{i}{2},\frac{1}{2},\frac{1}{2})$&$(-\frac{i}{2},-\frac{1}{2},\frac{1}{2})$&
$(-\frac{i}{2},\frac{1}{2},-\frac{1}{2})$&$(\frac{i}{2},-\frac{1}{2},-\frac{1}{2})$&
&&&&\\
&& 
$\stackrel{03}{\;\,}\;\;\,\stackrel{12}{\;\,}\;\;\,\stackrel{56}{\;\,}$&
$\stackrel{03}{\;\,}\;\;\,\stackrel{12}{\;\,}\;\;\,\stackrel{56}{\;\,}$&
$\stackrel{03}{\;\,}\;\;\,\stackrel{12}{\;\,}\;\;\,\stackrel{56}{\;\,}$&
$\stackrel{03}{\;\,}\;\;\,\stackrel{12}{\;\,}\;\;\,\stackrel{56}{\;\,}$&
&&&&\\
\hline
$$&$1$&$\stackrel{03}{(+i)}\stackrel{12}{(+)}\stackrel{56}{(+)}$ & 
$\stackrel{03}{[+i]}\stackrel{12}{[+]}\stackrel{56}{(+)}$ & 
$\stackrel{03}{[+i]}\stackrel{12}{(+)}\stackrel{56}{[+]}$ & 
$\stackrel{03}{(+i)}\stackrel{12}{[+]}\stackrel{56}{[+]}$&$\frac{i}{2}$&
$\frac{1}{2}$&$\frac{1}{2}$&$1$&$1$\\
$$&$2$&    $[-i][-](+) $ & $(-i)(-)(+) $ & $(-i)[-][+] $ & $[-i](-)[+] $ &$-\frac{i}{2}$&
$-\frac{1}{2}$&$\frac{1}{2}$&$1$&$1$\\
$$&$3$&    $[-i] (+)[-]$ & $(-i)[+][-] $ & $(-i)(+)(-) $ & $[-i][+](-) $&$-\frac{i}{2}$&
$\frac{1}{2}$&$-\frac{1}{2}$&$1$&$-1$ \\
$$&$4$&    $(+i)[-][-] $ & $[+i](-)[-] $ & $[+i][-](-) $ & $(+i)(-)(-) $&$\frac{i}{2}$&
$-\frac{1}{2}$&$-\frac{1}{2}$&$1$&$-1$ \\
\hline
$ odd\, II$&$$&$ $&$ $&$ $&
$ $&$S^{03}$&$S^{12}$&$S^{56}$&$\Gamma^{(5+1)}$&
$\Gamma^{(3+1)}$\\
&& 
$\stackrel{03}{\;\,}\;\;\,\stackrel{12}{\;\,}\;\;\,\stackrel{56}{\;\,}_{f_m}$&
$\stackrel{03}{\;\,}\;\;\,\stackrel{12}{\;\,}\;\;\,\stackrel{56}{\;\,}_{f_m}$&
$\stackrel{03}{\;\,}\;\;\,\stackrel{12}{\;\,}\;\;\,\stackrel{56}{\;\,}_{f_m}$&
$\stackrel{03}{\;\,}\;\;\,\stackrel{12}{\;\,}\;\;\,\stackrel{56}{\;\,}_{f_m}$&
&&&&\\
\hline
$$&$$&$(-i)(+)(+)_{d_4}$ & $[-i][+](+)_{d_3}$ & $[-i](+)[+]_{d_2}$ & $(-i)[+][+]_{d_1}$&
$-\frac{i}{2}$&$\frac{1}{2}$&$\frac{1}{2}$&$-1$&$-1$ \\
$$&$$&$[+i][-](+)_{c_4}$ & $(+i)(-)(+)_{c_3}$ & $(+i)[-][+]_{c_2}$ & $[+i](-)[+]_{c_1}$&
$\frac{i}{2}$&$-\frac{1}{2}$&$\frac{1}{2}$&$-1$&$-1$ \\
$$&$$&$[+i](+)[-]_{b_4}$ & $(+i)[+][-]_{b_3}$ & $(+i)(+)(-)_{b_2}$ & $[+i][+](-)_{b_1}$&
$\frac{i}{2}$&$\frac{1}{2}$&$-\frac{1}{2}$&$-1$&$1$ \\
$$&$$&$(-i)[-][-]_{a_4}$ & $[-i](-)[-]_{a_3}$ & $[-i][-](-)_{a_2}$ & $(-i)(-)(-)_{a_1}$&
$-\frac{i}{2}$&$-\frac{1}{2}$&$-\frac{1}{2}$&$-1$&$1$ \\
\hline\hline
$ even\, I$&$m$&$ $&$$&$ $&
$ $&$S^{03}$&$S^{12}$&$S^{56}$&$\Gamma^{(5+1)}$&
$\Gamma^{(3+1)}$\\
&&$(\frac{i}{2},\frac{1}{2},\frac{1}{2})$&$(-\frac{i}{2},-\frac{1}{2},\frac{1}{2})$&
$(\frac{i}{2},-\frac{1}{2},-\frac{1}{2})$&$(-\frac{i}{2},\frac{1}{2},-\frac{1}{2})$&
&&&&\\
&& 
$\stackrel{03}{\;\,}\;\;\,\stackrel{12}{\;\,}\;\;\,\stackrel{56}{\;\,}$&
$\stackrel{03}{\;\,}\;\;\,\stackrel{12}{\;\,}\;\;\,\stackrel{56}{\;\,}$&
$\stackrel{03}{\;\,}\;\;\,\stackrel{12}{\;\,}\;\;\,\stackrel{56}{\;\,}$&
$\stackrel{03}{\;\,}\;\;\,\stackrel{12}{\;\,}\;\;\,\stackrel{56}{\;\,}$&
&&&&\\
\hline
$$&$1$& $[-i](+)(+) $ & $(-i)[+](+) $ & $[-i][+][+] $ & $(-i)(+)[+] $ &$-\frac{i}{2}$&
$\frac{1}{2}$&$\frac{1}{2}$&$-1$&$-1$ \\ 
$$&$2$&    $(+i)[-](+) $ & $[+i](-)(+) $ & $(+i)(-)[+] $ & $[+i][-][+] $ &$\frac{i}{2}$&
$-\frac{1}{2}$&$\frac{1}{2}$&$-1$&$-1$ \\
$$&$3$&    $(+i)(+)[-] $ & $[+i][+][-] $ & $(+i)[+](-) $ & $[+i](+)(-) $&$\frac{i}{2}$&
$\frac{1}{2}$&$-\frac{1}{2}$&$-1$&$1$ \\
$$&$4$&    $[-i][-][-] $ & $(-i)(-)[-] $ & $[-i](-)(-) $ & $(-i)[-](-) $&$-\frac{i}{2}$&
$-\frac{1}{2}$&$-\frac{1}{2}$&$-1$&$1$ \\
\hline
$ even\, II$&$m$&$ $&$ $&$ $&
$ $&$S^{03}$&$S^{12}$&$S^{56}$&$\Gamma^{(5+1)}$&
$\Gamma^{(3+1)}$\\ 
&&$(-\frac{i}{2},\frac{1}{2},\frac{1}{2})$&$(\frac{i}{2},-\frac{1}{2},\frac{1}{2})$&
$(-\frac{i}{2},-\frac{1}{2},-\frac{1}{2})$&$(\frac{i}{2},\frac{1}{2},-\frac{1}{2})$&
&&&&\\
&& 
$\stackrel{03}{\;\,}\;\;\,\stackrel{12}{\;\,}\;\;\,\stackrel{56}{\;\,}$&
$\stackrel{03}{\;\,}\;\;\,\stackrel{12}{\;\,}\;\;\,\stackrel{56}{\;\,}$&
$\stackrel{03}{\;\,}\;\;\,\stackrel{12}{\;\,}\;\;\,\stackrel{56}{\;\,}$&
$\stackrel{03}{\;\,}\;\;\,\stackrel{12}{\;\,}\;\;\,\stackrel{56}{\;\,}$&
&&&&\\
\hline
$$&$1$&$[+i](+)(+) $ & $(+i)[+](+) $ & $[+i][+][+] $ & $(+i)(+)[+] $ &$\frac{i}{2}$&
$\frac{1}{2}$&$\frac{1}{2}$&$1$&$1$ \\
$$&$2$&$(-i)[-](+) $ & $[-i](-)(+) $ & $(-i)(-)[+] $ & $[-i][-][+] $ &$-\frac{i}{2}$&
$-\frac{1}{2}$&$\frac{1}{2}$&$1$&$1$ \\
$$&$3$&$(-i)(+)[-] $ & $[-i][+][-] $ & $(-i)[+](-) $ & $[-i](+)(-) $&$-\frac{i}{2}$&
$\frac{1}{2}$&$-\frac{1}{2}$&$1$&$-1$ \\
$$&$4$&$[+i][-][-] $ & $(+i)(-)[-] $ & $[+i](-)(-) $ & $(+i)[-](-) $&$\frac{i}{2}$&
$-\frac{1}{2}$&$-\frac{1}{2}$&$1$&$-1$ \\ 
\hline
 \end{tabular}
\end{tiny}
\end{table*}
%

Let us  find the solutions of the Weyl equation, Eq.~(\ref{Weyl}), taking into account 
four basis creation operators of the first family, $f=1(a)$, in Table~\ref{Table Clifffourplet.}. 
 Assuming that  moments in the fifth and the sixth dimensions are zero, 
$p^a=(p^0,p^1,p^2,p^3,0,0)$,  the following four plane wave solutions for positive energy, $p^0 = |\vec{p}|$, can be found, two with the positive charge $\frac{1}{2}$ 
and with spin $S^{12}$ either equal to $\frac{1}{2}$ or to  $-\frac{1}{2}$, and two  
with the negative charge $-\frac{1}{2}$ and again with  $S^{12}$ either  
$\frac{1}{2}$ or  $-\frac{1}{2}$.


%
\begin{eqnarray}
\label{weylgen0}
&& {\rm Clifford\; odd \;creation \; operators\;} {\rm in \;  d=(5+1)}
\, \nonumber\\
p^0&=&|p^0|\,, \;\; S^{56}= \frac{1}{2}\,, \;\;\Gamma^{(3+1)}=1\,,\nonumber\\
\Bigg( {\underline{\hat{\bf b}}}^{1 1 \dagger}_{tot} (\vec{p}) \, 
&=&
\beta\, \left( \stackrel{03}{(+i)}\,\stackrel{12}{(+)}| \stackrel{56}{(+)} + 
\frac{p^1 +i p^2}{ |p^0| + |p^3|} \stackrel{03}{[-i]}\,\stackrel{12}{[-]}|\stackrel{56}{(+)}
\right) \Bigg) \cdot \,\nonumber\\
&&e^{-i(|p^0| x^0 - \vec{p}\cdot\vec{x})}\,, \nonumber\\
\Bigg( {\underline{\hat{\bf b}}}^{2 1 \dagger}_{tot} (\vec{p}) \, 
&=& \beta^*\, \left(\stackrel{03}{[-i]}\,
\stackrel{12}{[-]}|\stackrel{56}{(+)} - \frac{p^1 -i p^2}{ |p^0| + |p^3|}\,
 \stackrel{03}{(+i)}\,\stackrel{12}{(+)}|\stackrel{56}{(+)}\right) \Bigg) \cdot\,\nonumber\\
&&e^{-i(|p^0| x^0 + \vec{p}\cdot\vec{x})}\,,\nonumber\\
&& {\rm Clifford\; odd \;creation \; operators\;} {\rm in \;  d=(5+1)}
\, \nonumber\\
p^0&=&|p^0|\,, \;\;S^{56}=- \frac{1}{2}\,,\;\;\Gamma^{(3+1)}=-1\,,\nonumber\\
\Bigg(  {\underline{\hat{\bf b}}}^{3 1 \dagger}_{tot} (\vec{p}) \,
&=& - \beta \, 
\left( \stackrel{03}{[-i]}\,\stackrel{12}{(+)}| \stackrel{56}{[-]} +
 \frac{p^1 +i p^2}{ |p^0| + |p^3|}
\stackrel{03}{(+i)}\,\stackrel{12}{[-]}|\stackrel{56}{[-]}\right) \Bigg) \cdot
\,\nonumber\\
&&e^{-i(|p^0| x^0 + \vec{p}\cdot\vec{x})}\,,\nonumber\\
\Bigg(  {\underline{\hat{\bf b}}}^{4 1 \dagger}_{tot} (\vec{p}) \,
&=& - \beta^*\,\left(\stackrel{03}{(+i)}
\,\stackrel{12}{[-]}| \stackrel{56}{[-]} - \frac{p^1 -i p^2}{ |p^0| + |p^3|} 
\stackrel{03}{[-i]}\,\stackrel{12}{(+)}|\stackrel{56}{[-]}\right) \Bigg) 
\cdot \,\nonumber\\
&&e^{-i(|p^0| x^0 - \vec{p}\cdot\vec{x})}\,,
%
\end{eqnarray}
%
%
Index ${}^{s=(1,2,3,4)}$ counts different solutions of the Weyl equations, 
index ${}^{f=1}$ denotes the family quantum number, all solutions belong to 
the same family, while 
$\beta^* \beta= \frac{|p^0| + |p^3|}{2|p^0|} $ takes 
care that the corresponding states are normalized. 

All four superposition of 
${\underline{\hat{\bf b}}}^{ s f \dagger}_{tot} (\vec{p})|_{p^0=|\vec{p}|}\, 
= \sum_{m} c^{ s f=1}{}_{m} (\vec{p},|p^0|=|\vec{p}|) \, \,
\hat{b}^{ m \dagger}_{f=1}\, e^{-i(p^0 x^0 -\varepsilon \vec{p}\cdot \vec{x})}$,
with $m=(1,2)$ for the 
first two states, and with $m=(3,4)$ for the second two states, 
Table~\ref{Table Clifffourplet.}, $s=(1,2,3,4)$, are orthogonal and 
correspondingly normalized, fulfilling Eq.~(\ref{Weylpp'solort}).









\end{small}


%
\section{Hilbert space of Clifford fermions}
\label{HilbertCliff0}

The Clifford odd creation operators ${\underline{\hat{\bf b}}}_{tot}^{s f \dagger} (\vec{p})$, with ${|p^0|=|\vec{p}|}$, are defined in Eq.~(\ref{creationb}) on the
tensor products of the $(2^{\frac{d}{2}-1})^2$ ''basis vectors'' (describing the
internal space of fermion fields) and of the (continuously) infinite number of basis 
in the momentum space. The solutions of the Weyl equation, Eq.~(\ref{Weyl}), are 
 plane waves  of particular momentum $\vec{p}$ and with energy related to momentum,
 ${|p^0|=|\vec{p}|}$.

 The creation operator 
${\underline{\hat{\bf b}}}_{tot}^{s f \dagger} (\vec{p})$ defines, when applied 
on the vacuum state $|\psi_{oc}>$, the $s^{th}$ of the $2^{\frac{d}{2}-1}$ plane
wave solutions of a particular momentum $\vec{p }$ belonging to the $f^{th}$ of
the $2^{\frac{d}{2}-1}$ ''families''. They  fulfill together with the Hermitian
conjugated partners annihilation operators  
${\underline{\hat{\bf b}}}_{tot}^{s f} (\vec{p})$ 
 the anticommutation relations of Eq.~(\ref{Weylpp'comrel}). 
 
 These creation operators form the Hilbert space of ''Slater determinants'', defining for
 each ''Slater determinant'' the ''space'' for any of the single particle fermion states of an odd Clifford character, due to the oddness of the ''basis vector'' of an odd Clifford 
 character.
 Each of these ''spaces'' can be empty or occupied. Correspondingly there is the 
''Slater determinant'' with all the ''spaces'' empty, the ''Slater determinants'' with only
one of the ''spaces'' occupied, any one, and all the rest empty, the ''Slater determinants'' with two ''spaces'' occupied, any two, and all the rest empty, and so on. 

These ''Slater determinant'' of all possible occupied and empty states can be 
explained  as well if introducing  the tensor multiplication of single fermion 
states of any quantum number and any momentum, with the constant included.

{\bf  Statement 3}: Introducing the tensor product multiplication $*_{T}$ of any 
number of  Clifford odd fermion states of all possible internal quantum 
numbers and all possible momenta (that is of any number of 
$  \underline{\hat {\bf b}}^{s \,f\, \dagger}_{tot} (\vec{p})$) of any
 $(s,f, \vec{p})$ we generate the Hilbert space of Clifford fermions.

The Hilbert space of a particular momentum $\vec{p}$, ${\cal H}_{\vec{p}}$, 
contains the finite number of ''Slater determinants''. The number of
 ''Slater determinants'' is in $d$-dimensional space equal to
\begin{eqnarray}
N_{{\cal H}_{\vec{p}}}& =&  2^{2^{d-2}}\,.
\label{NHp}
\end{eqnarray}
The total Hilbert space of  anticommuting fermions is the product  $\otimes_{N}$
of the Hilbert spaces of particular $\vec{p}$
\begin{eqnarray}
{\cal H}& =& \prod_{\vec{p}}^{\infty}\otimes_{N} {\cal H}_{\vec{p}}\,.
\label{H}
\end{eqnarray}
The total Hilbert space ${\cal H}$ is correspondingly infinite and  contains
$N_{\cal H}$ ''Slater determinants''
\begin{eqnarray}
N_{\cal H}& =& \prod_{\vec{p}}^{\infty} 2^{2^{d-2}}\,.
\label{NH}
\end{eqnarray}
%

Before starting to comment the application of the creation operators  
${\underline{\hat{\bf b}}}_{tot}^{s f \dagger} (\vec{p})$ and annihilation
${\underline{\hat{\bf b}}}_{tot}^{s f} (\vec{p})$ operators on the Hilbert space
${\cal H}$ (described with all possible ''Slater determinants'' of all possible 
occupied and empty fermion states of all possible $(s,f,\vec{p})$, or by the tensor 
products of all possible single fermion states of all possible $(s,f,\vec{p})$, with 
the identity included) 
let us discuss properties of creation and annihilation operators, the anticommutation relations of which are presented in Eq.~(\ref{Weylpp'comrel}).
%

The creation operators $\underline{\hat {\bf b}}^{s f \dagger}_{tot} (\vec{p})$ and 
the annihilation operators $\underline{\hat {\bf b}}^{s' f' }_{tot} (\vec{p'})$, having an odd Clifford character, anticommute, manifesting the properties as follows
%
\begin{eqnarray}
\label{tensorproperties}
\underline{\hat {\bf b}}^{ s f \dagger }_{tot} (\vec{p})*_{T} 
\underline{\hat {\bf b}}^{s' f' \dagger}_{tot} (\vec{p}{\,}')&=& - 
\underline{\hat {\bf b}}^{s' f' \dagger}_{tot} (\vec{p}{\,}')*_{T} 
\underline{\hat {\bf b}}^{s f \dagger}_{tot} (\vec{p})\,, \nonumber\\
\underline{\hat {\bf b}}^{ s f }_{tot} (\vec{p})*_{T} 
\underline{\hat {\bf b}}^{s' f' }_{tot} (\vec{p}{\,}')&=& - 
\underline{\hat {\bf b}}^{s' f' }_{tot} (\vec{p}{\,}')*_{T} 
\underline{\hat {\bf b}}^{ s f }_{tot} (\vec{p})\,, \nonumber\\
\underline{\hat {\bf b}}^{ s f }_{tot} (\vec{p})*_{T} 
\underline{\hat {\bf b}}^{s' f' \dagger}_{tot} (\vec{p}{\,}')&=& - 
\underline{\hat {\bf b}}^{s' f' \dagger}_{tot} (\vec{p}{\,}')*_{T} 
\underline{\hat {\bf b}}^{ s f }_{tot} (\vec{p})\,, \nonumber\\ 
 {\rm if \;\, at \;\, least \,\; one \,\; of \,} (s,f, \vec{p}) && {\rm is \; different\,\;from\;} 
(s',f', \vec{p}{\,}') \,,\nonumber\\
\underline{\hat {\bf b}}^{ s f \dagger }_{tot} (\vec{p})*_{T} 
\underline{\hat {\bf b}}^{ s f \dagger }_{tot} (\vec{p})&=& 0\,,\nonumber\\
\underline{\hat {\bf b}}^{ s f }_{tot} (\vec{p})*_{T} 
\underline{\hat {\bf b}}^{ s f }_{tot} (\vec{p}) &=& 0\,,\nonumber\\ 
\underline{\hat {\bf b}}^{ s f }_{tot} (\vec{p})*_{T}
\underline{\hat {\bf b}}^{ s f \dagger}_{tot} (\vec{p}) &=& 1\, ({\rm identity})
\,,\nonumber\\
\underline{\hat {\bf b}}^{s f }_{tot} (\vec{p}) |\psi_{oc}>&=& 0\,.
\end{eqnarray}
The above relations, leading from the commutation relations of 
Eq.~(\ref{Weylpp'comrel}),  determine the rules of the application of  creation and 
annihilation operators on  ''Slater determinants'': \\
{\bf i.} The creation operator 
$\underline{\hat {\bf b}}^{ s f \dagger}_{tot} (\vec{p})$
jumps over the creation operators determining the occupied state of another kind (that is over the occupied state distinguishing from the jumping creation one in any of the 
internal quantum numbers ($s,f$) or in $\vec{p}$) up to the last step when it comes 
to its own empty state with the quantum numbers ($f,s$) and $\vec{p}$,  
occupying this empty state, or, if this state is already occupied, gives zero.  Whenever  
$\underline{\hat {\bf b}}^{ s f \dagger}_{tot} (\vec{p})$ jumps over an occupied state  changes the sign of the ''Slater determinant''.\\
{\bf ii.} The annihilation operator changes the sign whenever jumping over the occupied state carrying different internal quantum numbers ($s,f$) or  $\vec{p}$, unless it comes to the occupied state with its own internal quantum numbers  ($s,f$) and its own $\vec{p}$, emptying this state, or, if this state is empty, gives zero.

 Let us point out that the Clifford odd  creation operators,  
$\underline{\hat {\bf b}}^{ s f \dagger}_{tot} (\vec{p})$, and annihilation operators, 
$\underline{\hat {\bf b}}^{ s' f'}_{tot} (\vec{p'})$, fulfill the anticommutation 
relations of Eq.~(\ref{Weylpp'comrel}) for any  $\vec{p}$ and any  $(s,f)$ due to the anticommuting character (the Clifford oddness)  of the ''basis vectors'', 
$\hat{b}^{m \dagger}_{f}$ and their Hermitian conjugated partners 
$\hat{b}_{f}^{m}$, Eqs.~(\ref{start(2n+1)2cliffgammatilde4n}, \ref{d=2(2n+1)}), 
what means that  the anticommuting character  of creation and annihilation operators 
is not postulated.  

The total $\,$ Hilbert space ${\cal H}$ has infinite  number of degrees of freedom 
(of ''Slater determinants'') due to the infinite number of Hilbert spaces 
${\cal H}_{\vec{p}}$ of particular $\vec{p}$, ${\cal H} $ 
$= \prod_{\vec{p}}^{\infty}\otimes_{N} {\cal H}_{\vec{p}}$, while the Hilbert 
space ${\cal H}_{\vec{p}}$ of particular momentum ${\vec{p}}$ has the finite 
dimension $2^{2^{d-2}}$.

In Subsects.~\ref{HilbertCliffp}, \ref{HilbertCliffgen}, \ref{IllustrationH}
 the properties of Hilbert spaces are discussed in more details.

%


%
\subsection{Application of ${\underline{\hat{\bf b}}}_{tot}^{s f \dagger}
 (\vec{p})$ and ${\underline{\hat{\bf b}}}_{tot}^{s f} (\vec{p})$ on Hilbert 
 space of Clifford fermions of particular $\vec{p}$}
\label{HilbertCliffp}

The $2^{d-2}$ Clifford odd creation operators  of particular momentum 
$\vec{p }$, $\underline{\hat{\bf b}}_{tot}^{s f \dagger} (\vec{p}, p^0)$, 
with the property $|p^0|=|\vec{p }|$, each representing the $s^{th}$ 
solution of Eq.~(\ref{Weyl}) for a particular family $f$, fulfill  together 
with the (Hermitian conjugated partners) annihilation operators 
$\underline{\hat{\bf b}}_{tot}^{s f} (\vec{p})$ the anticommutation 
relations of Eq.~(\ref{Weylpp'comrel}), the application of which on the 
Hilbert space of ''Slater determinants'' are discussed in  
Eq.~(\ref{tensorproperties}) and in the text below this equation.

The Hilbert space  ${\cal H}_{\vec{p}}$ of a particular momentum $\vec{p}$
 consists correspondingly of $2^{2^{d-2}}$ ``Slater determinants''. 
%
%
Let us write down explicitly these $2^{2^{d-2}}$ contributions to the Hilbert space 
${\cal H}_{\vec{p}}$  of a particular momentum ${\vec{p}}$, using the notation that 
${\bf 0^{s f}}_{\vec{p}}$ represents the unoccupied state 
$|\psi^{s f} (\vec{p}, p^0)>~|_{|p^0|=|\vec{p }|}=$ 
$\underline{\hat{\bf b}}_{tot}^{s f \dagger} (\vec{p})|_{|p^0|=|\vec{p}|}\,
|\psi_{oc}>$ of the $s^{th}$ solution of the equations of motion for the $f^{th}$ family and the momentum $|p^0|=|\vec{p}|$),  Eq.~(\ref{Weylp}),
while ${\bf 1^{s f}}_{\vec{p}}$  represents the corresponding occupied state.

 The number operator is defined as
%
 \begin{eqnarray}
\label{NOSD}
N^{s f}_{\vec{p}}&=& {\underline{\hat{\bf b}}}_{tot}^{s f \dagger} (\vec{p})\,*_{T}\,
 \underline{\hat{\bf b}}_{tot}^{s f} (\vec{p})\,,\nonumber\\
N^{s f}_{\vec{p}}\, |\psi_{oc}>&=& 0\cdot |\psi_{oc}>\,,\quad\;
\quad N^{s f}_{ \vec{p}}\, *_{T}\, {\bf 0^{s f}}_{\vec{p}}=0\,,\nonumber\\
N^{s f}_{ \vec{p}} \, *_{T}\,{\bf 1^{s f}}_{\vec{p}}&=&
1\,\cdot {\bf 1^{s f}}_{\vec{p}} \,, \quad
N^{s f}_{ \vec{p}} \,*_{T}\, N^{s f}_{ \vec{p}} \,*_{T} {\bf 1^{s f}}_{\vec{p}}\,=
1\cdot {\bf 1^{s f}}_{\vec{p}} \,.
\end{eqnarray}
One can  check the above relations on the example of $d=(5+1)$, with the
"basis vectors" for $f=1$ presented in 
Table~\ref{Table Clifffourplet1.} and with the solution for Weyl equation, Eq.~(\ref{Weyl}),
presented in Eq.~(\ref{weylgen0}). 
\begin{table}
\begin{tiny}     
\begin{center}
\caption{\label{Table Clifffourplet1.} 
 The four creation operators of the  irreducible representation 
{\it odd I} from Table~\ref{Table Clifffourplet.}, $d=(5+1)$, $f=1(a)$. together with their 
Hermitian conjugated partners are presented (up to a phase). 
}
  \begin{tabular}{|c|c|c|}
\hline
$i$&$ f=1(a)$&$\rm{Her. \,con.}$$\,f=1(a)$\\
\hline
$1$&$\stackrel{03}{(+i)}\stackrel{12}{(+)}\stackrel{56}{(+)}$ & 
$\stackrel{03}{(-i)}\stackrel{12}{(-)}\stackrel{56}{(-)}$\\
$2$&    $\stackrel{03}{[-i]}\stackrel{12}{[-]}\stackrel{56}{(+)} $ &
                $\stackrel{03}{[-i]}\stackrel{12}{[-]}\stackrel{56}{(-)} $ \\
$3$&    $\stackrel{03}{[-i]}\stackrel{12}{(+)}\stackrel{56}{[-]}$ & 
                $\stackrel{03}{[-i]}\stackrel{12}{(-)}\stackrel{56}{[-]} $\\
$4$&    $\stackrel{03}{(+i)}\stackrel{12}{[-]}\stackrel{56}{[-]} $ & 
                $\stackrel{03}{(-i)}\stackrel{12}{[-]}\stackrel{56}{[-]} $ \\
\hline
 \end{tabular}
\end{center}
\end{tiny}
\end{table}
%


Let us write down the Hilbert space of second quantized fermions ${\cal H}_{\vec{p}}$,
using the simplified  notation as in Part I, Sect. III.A., counting for   $f=1$  empty states 
as ${\bf 0_{r p}}$, and occupied states as ${\bf 1_{r p}}$, with 
$r=(1,\dots,  2^{\frac{d}{2}-1})$, 
for $f=2$ we count $r= 2^{\frac{d}{2}-1} +1,\cdots,  2^{d-2}$.
Correspondingly we can represent  ${\cal H}_{\vec{p}}$ as follows
\begin{eqnarray}
\label{SD}
|{\bf 0_{1 p}}, {\bf 0_{2 p}}, {\bf 0_{3 p}}, \dots, {\bf 0_{2^{d-2} p}}>|_{1}&&,\nonumber\\
|{\bf 1_{1 p}}, {\bf 0_{2 p}}, {\bf 0_{3 p}}, \dots, {\bf 0_{2^{d-2} p}}>|_{2}&&,\nonumber\\
|{\bf 0_{1 p}}, {\bf 1_{2 p}}, {\bf 0_{3 p}}, \dots, {\bf 0_{2^{d-2} p}}>|_{3}&&,\nonumber\\
|{\bf 0_{1 p}}, {\bf 0_{2 p}}, {\bf 1_{3 p}}, \dots, {\bf 0_{2^{d-2} p}}>|_{4}&&,\nonumber\\
\vdots &&\nonumber\\
|{\bf 1_{1 p}}, {\bf 1_{2 p}}, {\bf 0_{3 p}}, \dots, {\bf 0_{2^{d-2} p}}>|_{2^{d-2} +2}&&,\nonumber\\%
\vdots &&\nonumber\\|{\bf 1_{1 p}}, {\bf 1_{2 p}}, {\bf 1_{3 p}}, \dots, {\bf 1_{2^{d-2} p}}>|_{2^{2^{d-2}} }&&\,,
\end{eqnarray}
with a part with none of states occupied ($N_{r p} =0$ for all $r=1,\dots, 2^{d-2}$),
with a part with only one of states occupied ($N_{r p}=1$ for a particular 
$r=(1,\dots, 2^{d-2})$, while  $N_{r' p}=0$ for all the others  $r' \ne r$),
with a part with only two of states occupied ($N_{r p}=1$ and $N_{r' p}=1$, where
$r$ and $r'$ run from $(1,\dots, 2^{d-2}$), and so on. The last part has all the states
occupied.

It is not difficult to see that the creation and annihilation operators, when applied on this
Hilbert space ${\cal H}_{\vec{p}}$, fulfill the anticommutation relations for the second
quantized Clifford fermions.
\begin{eqnarray}
\{\underline{\hat{\bf b}}_{tot}^{s f} (\vec{p})\,, \underline{\hat{\bf b}}_{tot}^{s' f' \dagger} (\vec{p})\}_{*_{T}+} 
{\cal H}_{\vec{p}} &=& \delta^{s s'}\; \delta^{f f'} 
{\cal H}_{\vec{p}}\,,\nonumber\\
\{\underline{\hat{\bf b}}_{tot}^{s f} (\vec{p}), \underline{\hat{\bf b}}_{tot}^{s' f'} (\vec{p}) \}_{*_{T}+}\;  
{\cal H}_{\vec{p}}&=& 0\;\cdot {\cal H}_{\vec{p}} \,,\nonumber\\
\{\underline{\hat{\bf b}}_{tot}^{s f \dagger} (\vec{p}) \,,
\underline{\hat{\bf b}}_{tot}^{s' f' \dagger} (\vec{p})\}_{*_{T}+}\; {\cal H}_{\vec{p}}&=&
0\;\cdot {\cal H}_{\vec{p}}\,.
\label{CliffcomrelHp}
\end{eqnarray}
The proof for the above relations easily follows if one takes into account that 
whenever the creation or annihilation operator jumps over an odd products of 
occupied states the sign of the ''Slater determinant'' changes due to the 
oddness of the occupied states, while states, belonging to different $\vec{p}$ are 
orthogonal~\footnote{The orthogonality of the states are even easier to be 
visualized since the two delta functions at $\vec{x}$ and at $\vec{x}{\,}'$,
$\vec{x}\ne \vec{x}{\,}'$ are obviously orthogonal.}, see 
Eq.~(\ref{tensorproperties}) and  the text below this equation.
Then one sees that the contribution of the application of 
$\underline{\hat{\bf b}}_{tot}^{s f \dagger} (\vec{p}) \,*_{T}\,$
$\underline{\hat{\bf b}}_{tot}^{s' f'} (\vec{p})\;*_{T}\,$ on 
$ {\cal H}_{\vec{p}}$ has the opposite sign than the contribution of  
$\underline{\hat{\bf b}}_{tot}^{s' f'} (\vec{p})$
$\,*_{T}\,\underline{\hat{\bf b}}_{tot}^{s f \dagger} (\vec{p})\; *_{T}\,$ on
${\cal H}_{\vec{p}}$. 

If the creation and annihilation operators are Hermitian conjugated to each other, 
the result follows
\[(\,\underline{\hat{\bf b}}_{tot}^{s f} (\vec{p})\,*_{T}\,
\underline{\hat{\bf b}}_{tot}^{s f \dagger} (\vec{p}) +
\underline{\hat{\bf b}}_{tot}^{s f \dagger} (\vec{p})\, *_{T}\,
\underline{\hat{\bf b}}_{tot}^{s f \,} (\vec{p}) \,)\,*_{T}\, {\cal H}_{\vec{p}}= 
{\cal H}_{\vec{p}}\,,\] manifesting that this application of ${\cal H}_{\vec{p}}$
 gives the whole ${\cal H}_{\vec{p}}$ back. Each of the two summands operates 
 on their own half of ${\cal H}_{\vec{p}}$. Jumping together over an even number 
 of occupied states,  $\underline{\hat{\bf b}}_{tot}^{s f } (\vec{p})$ and 
 ${\hat{\bf b}}_{tot}^{s f \dagger} (\vec{p})$ do 
 not change the sign of the particular ``Slater determinant''.  
(Let us add that  $\underline{\hat{\bf b}}_{tot}^{s f}  (\vec{p})$  reduces for the 
particular $s$ and $f$ the Hilbert space ${\cal H}_{\vec{p}}$ for the factor  
$\frac{1}{2}$, and so does 
$\underline{\hat{\bf b}}_{tot}^{s f \dagger} (\vec{p})$. 
The sum of both, applied on ${\cal H}_{\vec{p}}$, reproduces the whole 
${\cal H}_{\vec{p}}$.) 

Let us repeat that the number of ''Slater determinants'' in the Hilbert space of particular 
momentum $\vec{p}$, ${\cal H}_{\vec{p}}$, in $d$-dimensional space is finite  and equal to
$N_{{\cal H}_{\vec{p}}} =  2^{2^{d-2}}$\,.
%


%
\subsection{Application of ${\underline{\hat{\bf b}}}_{tot}^{s f \dagger}
 (\vec{p})$ and ${\underline{\hat{\bf b}}}_{tot}^{s f} (\vec{p})$ on total Hilbert space 
 ${\cal H}$ of Clifford fermions}
\label{HilbertCliffgen}

The total Hilbert space of  anticommuting fermions is the infinite product of the Hilbert spaces of particular $\vec{p}$, Eq.~(\ref{H}), 
%
$ {\cal H}= \prod_{\vec{p}}^{\infty}\otimes_{N} {\cal H}_{\vec{p}}\,$.
%

Due to the Clifford odd character of creation and annihilation operators, 
Eq.~(\ref{Weylpp'comrel}), and the orthogonality of the plane waves belonging to different 
momenta $\vec{p}\,$,  it follows that $\underline{\hat{\bf b}}_{tot}^{s f \dagger} (\vec{p})$ 
$*_{T}\,\underline{\hat{\bf b}}_{tot}^{s f \dagger} (\vec{p}{\,}') \,*_{T}\,{\cal H} \ne 0 $, $\vec{p}\ne \vec{p}{\,}'$,  while $\{\,\underline{\hat{\bf b}}_{tot}^{s f \dagger} (\vec{p})\,*_{T}\, $ 
 $\underline{\hat{\bf b}}_{tot}^{s f \dagger} (\vec{p}{\,}') +$ 
$\underline{\hat{\bf b}}_{tot}^{s f \dagger} (\vec{p}{\,}')\, *_{T}\, $ 
$\underline{\hat{\bf b}}_{tot}^{s f \dagger} (\vec{p})\,\}\,*_{T}\, {\cal H} =0$, 
$\vec{p}\ne \vec{p}{\,}'$. This can be proven if taking into account Eq.~(\ref{tensorproperties}).  
For ``plane wave solutions'' of equations of motion in a box the momentum $\vec{p}$
 is discretized, otherwise  is continuous. The number of ``Slater determinants'' in the Hilbert space ${\cal H}$ in $d$-dimensional space is infinite (in both cases)
%
$N_{\cal H} = \prod_{\vec{p}}^{\infty} 2^{2^{d-2}}\,$.
%

Since the creation operators $\underline{\hat{\bf b}}_{tot}^{s f \dagger} 
(\vec{p})$ and the annihilation operators 
$\underline{\hat{\bf b}}_{tot}^{s' f'} (\vec{p}{\,}')$ fulfill for particular 
$\vec{p}$ the anticommutation relations on ${\cal H}_{\vec{p}}$, 
Eq.~(\ref{CliffcomrelHp}),  and since the momentum states, the 
plane wave solutions, are 
orthogonal, and correspondingly the creation and annihilation operators defined
on the tensor products of the internal basis and the momentum basis, representing
 fermions, anticommute, Eq.~(\ref{Weylpp'comrel}) (the Clifford odd objects 
$\underline{\hat{\bf b}}_{tot}^{s f \dagger} (\vec{p})$ demonstrate their 
oddness also with respect to 
$\underline{\hat{\bf b}}_{tot}^{s f \dagger} (\vec{p}{\,}')$), 
the anticommutation relations follow also for the application of 
$\underline{\hat{\bf b}}_{tot}^{s f \dagger} (\vec{p})$ and 
$\underline{\hat{\bf b}}_{tot}^{s f } (\vec{p})$
on ${\cal H}$
\begin{eqnarray}
\{\underline{\hat{\bf b}}_{tot}^{s f} (\vec{p}) \,, 
\underline{\hat{\bf b}}_{tot}^{s' f' \dagger} (\vec{p}{\,}') \}_{*_{T}+} \,{\cal H} &=& \delta^{s s'}\; \delta_{f f'}\; \delta (\vec{p} -\vec{p}{\,}')\;{\cal H}\,,\nonumber\\
\{\underline{\hat{\bf b}}_{tot}^{s f \dagger} (\vec{p}),{\hat{\bf b}}_{tot}^{s' f' \dagger} 
(\vec{p}{\,}')\}_{*_{T}+}\; {\cal H}&=& 0\;\cdot {\cal H} \,,\nonumber\\
\{\underline{\hat{\bf b}}_{tot}^{s f \dagger} (\vec{p}),
{\hat{\bf b}}_{tot}^{s' f' \dagger} (\vec{p}{\,}') \}_{*_{T}+}\; {\cal H}&=&
0\;\cdot{\cal H}\,.
\label{ijthetaprodgenH}
\end{eqnarray}
%
%


%
\subsection{Illustration of ${\cal H}$ in $d=(1+1)$}
\label{IllustrationH}
\begin{small}

Let us illustrate the properties of 
${\cal H}$ and the application of the creation operators on ${\cal H}$ in $d=(1+1)$ dimensional space in a toy model with two discrete 
momenta ($p^1_1, p^1_2$). Generalization to many momenta is straightforward.

The internal space of fermions contains only one creation operator, one ``basis vector'' 
$\hat{b}^1_{1}$ $=\stackrel{01}{(+i)}$, one family member $m=1$
 of the only family $f=1$. 
Correspondingly the creation operators  $\underline{\hat{\bf b}}^{1  1\dagger}_{tot} (\vec{p^1_i})|_{|p^0|=|p^1_{i}|} \, {\bf :}\,$  
$=\stackrel{01}{(+i)} \,e^{-i(p^0 x^0 -p^1_{i} x^1)}|_{|p^1_i| =|p^0_i| }$  
 distinguish only in momentum space of the fermion degrees of freedom. Their Hermitian 
conjugated annihilation operators are 
$\underline{\hat{\bf b}}^{11}_{tot} (\vec{p^1 _i})_{|p^0|=|p^1_{i}|}$, while the vacuum state is  $|\psi_{oc}>$  $=\stackrel{01}{(-i)}\cdot
\stackrel{01}{(+i)}=\stackrel{01}{[-i]} $. 

The whole Hilbert space for this toy model has correspondingly four "Slater determinants", numerated by
$|\quad>_{i}, i=(1,2,3,4)$
\begin{eqnarray}
(|{\bf 0_{p_1}} {\bf 0_{p_2}}>|_1\,, \, |{\bf 1_{p_1}} {\bf 0_{p_2}}>|_2\,, \,
|{\bf 0_{p_1}} {\bf 1_{p_2}}>|_3\,, \, |{\bf 1_{p_1}} {\bf 1_{p_2}}>|_4)\,,\nonumber
\end{eqnarray}
${\bf 0_{p^1_i}}$ represents an empty state and ${\bf 1_{p^1_i}}$ the occupied state.
Let us evaluate the application of $\{\underline{\hat{\bf b}}^{1  1}_{tot} (\vec{p^1_1})\,,$ 
$\underline{\hat{\bf b}}^{1  1\dagger}_{tot} (\vec{p^1_2})\}_{*_{T}+}$ on the
Hilbert space ${\cal H}$.
It follows
\begin{eqnarray}
&&\{\underline{\hat{\bf b}}^{1  1}_{tot} (\vec{p}^1_1)\,, 
\underline{\hat{\bf b}}^{1  1\dagger}_{tot} (\vec{p}^1_2)\}_{*_{T}+} {\cal H}=\nonumber\\
&&\underline{\hat{\bf b}}^{1  1}_{tot} (\vec{p}^1_1)\,*_{T}\, 
(|{\bf 0_{p_1}} {\bf 1_{p_2}}>|_{1\to 3} \,,\,
-|{\bf 1_{p_1}} {\bf 1_{p_2}}>|_{2\to 4}) + \nonumber\\
&&\underline{\hat{\bf b}}^{1  1\dagger}_{tot} (\vec{p}^1_2)\, *_{T}\, 
( |{\bf 0_{p_1}} {\bf 0_{p_2}}>_{2\to 1}\,, \, 
 +  |{\bf 0_{p_1}} {\bf 1_{p_2}}>_{4\to3})=\nonumber\\
&& (-|{\bf 0_{p_1}} {\bf 1_{p_2}>}_{2\to 4 \to 3}+
|{\bf 0_{p_1}} {\bf 1_{p_2}}>_{2\to 1\to 3})\,=0\,. \nonumber
\end{eqnarray}
\end{small}

%

%
\subsection{Relation between second quantized fermions of Dirac and second quantized 
fermions originated in odd Clifford algebra }
\label{bethenormarelation}

The Clifford odd creation operators 
$\underline{\hat{\bf b}}_{tot}^{s f \dagger} (\vec{p}) $ and their 
Hermitian conjugated partners annihilation operators 
$\underline{\hat{\bf b}}_{tot}^{s f} (\vec{p}) $ obey the
anticommutation relations of Eq.~(\ref{ijthetaprodgenH}) --- on the vacuum state 
$|\psi_{oc}>$, Eq.~(\ref{vac1}), and on the  whole Hilbert space ${\cal H}$,
Eq.~(\ref{ijthetaprodgenH}).
Creation operators, $\underline{\hat{\bf b}}_{tot}^{s f \dagger} (\vec{p})$,
 operating on a vacuum state, as well as on the whole Hilbert space, define second quantized fermion states. 
 
%
 

Let us relate here the Dirac's second quantization relations and the relations between 
creation operators  $\underline{\hat{\bf b}}_{tot}^{s f} (\vec{p}) $ and their Hermitian conjugated partners annihilation operators, without paying attention on the charges and family quantum numbers, since Dirac's creation operators do not pay attention on these two kinds of quantum numbers. We shall relate vectors in $d=(3+1)$ of both origins. 

In the Dirac case the second quantized field operators  are in $d=(3+1)$ dimensions 
postulated as follows
\begin{eqnarray}
\label{stateDirac}
{\underline {\bf {\Huge \Psi}}}^{h s\dagger}(\vec{x}, x^0)& =&\sum_{m, \vec{p}_k} \hat{{\bf a}}^{h \dagger}_m (\vec{p}_k)\, 
v^{h s}_m (\vec{p}_k) \,. 
\end {eqnarray}
$v^{h s}_m (\vec{p}_k)= u^{h s}_m (\vec{p_k})$ 
$ e^{-i(p^0 x^0- \varepsilon \vec{p}_k \cdot \vec{x})}$ are the two  left handed  
($\Gamma^{(3+1)}=-1=h$) and  the two right handed ($\Gamma^{(3+1)}= 1=h$, 
Eq.~(B.3)) two-component column matrices, $m=(1,2)$, 
representing the twice two solutions $s$ of the Weyl equation for free massless fermions 
of particular momentum $|\vec{p}_k|= |p_{k}^{0}|$~(\cite{BetheJackiw},
 Eqs.~(20-49) - (20-51)), the factor $\varepsilon =\pm1$ depends on the product of 
 handedness and spin.

$\hat{{\bf a}}^{h \dagger}_m (\vec{p}_k)$ are by Dirac postulated creation operators, 
which together  with the annihilation operators $\hat{{\bf a}}^{h}_m (\vec{p}_k)$, fulfill the 
anticommutation relations~(\cite{BetheJackiw}, Eqs.~(20-49) - (20-51))
 \begin{eqnarray}
 \label{comDirac}
\{\hat{{\bf a}}^{h\dagger}_m (\vec{p}_k), \,\hat{{\bf a}}^{h'\dagger}_n (\vec{p}_l)\}_{*_{T}+}&=&
0= \{\hat{{\bf a}}^{h}_m (\vec{p}_k), \,\hat{{\bf a}}^{h'}_n (\vec{p}_l)\}_{*_{T}+}\,,\nonumber\\ 
\{\hat{{\bf a}}^{h}_m (\vec{p}_k), \,\hat{{\bf a}}^{h'\dagger}_n (\vec{p}_l)\}_{*_{T}+} &=&
\delta_{mn} \delta^{h h'} \delta_{\vec{p}_k \vec{p}_l}\,
\end{eqnarray}
in the case of discretized  momenta for a fermion in a box. (Massive fermions are represented by four vectors which are the superposition of both handedness.)

Let us present the  two ''basis vectors''  $\hat{b}^{h \dagger}_m, m=(1,2), h $
representing left and right handedness, in the internal space of fermions 
in $d=(3+1)$, described by the Clifford odd algebra, representing the creation 
operators of one particular family ($f$ not shown in this case), without charges, 
of one handedness and with spins $\pm \frac{1}{2}$, respectively, operating 
on the vacuum state $|\psi_{oc}>=$ $\stackrel{03}{[+i]}\stackrel{12}{[-]}${\bf :} 
 $\hat{b}^{h \dagger}_1 =\stackrel{03}{[+i]}\stackrel{12}{(+)}$ and 
$\hat{b}^{h \dagger}_2=\stackrel{03}{(-i)}\stackrel{12}{[-]}$, 
Eq.~(\ref{start(2n+1)2cliffgammatilde4n}, \ref{d=2(2n+1)})~%
\footnote{We choose in the Clifford case the first two members of the third family 
in Table~\ref{Table Clifffourplet.}, since they manifest in $d=(3+1)$ the Clifford odd character.}, with $h=1$, representing the right handedness. %
These two ''basis vectors''  should be compared with the two vectors, one 
corresponding to the spin $\frac{1}{2}$ and the other to the spin 
$- \frac{1}{2}$ in the Dirac case.

Since Dirac did not postulate such creation operators on the level of 
$\hat{b}^{h \dagger}_m$,  let us postulate them now on the level
of $\hat{b}^{h \dagger}_m$, to be able to compare 
in this paper presented creation operators for this particular case, 
${\hat a}^{h \dagger}_{\uparrow}$ 
and $ {\hat a}^{h\dagger}_{\downarrow}$, of right handedness $h$ and spin up and 
down ($\uparrow, \downarrow$)
as follows
\begin{displaymath}
\hat{b}^{h \dagger}_{1} = \stackrel{03}{[+i])}\stackrel{12}{(+)}\,\quad 
{\rm to\; be\; related\; to}\quad  {\hat a}^{h \dagger}_{\uparrow}\,,\nonumber\\
\qquad \hat{b}^{h \dagger}_{2} = \stackrel{03}{(-i))}\stackrel{12}{[-]}\,\quad 
{\rm to\; be\; related\; to}\quad  {\hat a}^{h \dagger}_{\downarrow}\,.
 \end{displaymath}
One should repeat this also for left handedness $h=-1$. But these creation operators 
${\hat a}^{h \dagger}_{m}, m=(1,2)=(\uparrow, \downarrow)$, still can not be 
compared with the Dirac's ones.

Let us make the superposition of both  creation operators of particular handedness $h$, 
${\bf {\hat a}}^{h s \dagger} (\vec{p}_k):=$
$\alpha^{h s}_{\uparrow} (\vec{p}_{k}) \, {\hat a}^{h \dagger}_{\uparrow} +$
$ \alpha^{h s}_{\downarrow} (\vec{p}_{k}) \,{\hat a}^{h \dagger}_{\downarrow},
$ with the coefficients $\alpha^{hs}_{\uparrow} (\vec{p}_{k}) $ and 
$\alpha^{hs}_{\downarrow} (\vec{p}_{k}) $ chosen so that  
$\underline{\bf {\hat a}}^{h s \dagger}_{tot} (\vec{p}_k) {\bf :}=
{\bf {\hat a}}^{h s \dagger} (\vec{p}_k) \,e^{-i(p^0 x^0-\vec{p}_{k}\cdot \vec{x})}$
solves the equations of motion, Eq.~(\ref{Weyl})~\footnote{
 The equations of motion read in the Dirac case: 
 $\{\hat{p}^0 + (-2i S^{0i}\hat{p}_i)\}
  (\alpha^{s}_{1} (\vec{p}_{k}) \, {\bf {\hat a}}^{\dagger}_{1}$
$+ \alpha^{s}_{2} (\vec{p}_{k}) {\,\bf {\hat a}}^{\dagger}_{2})
e^{-i(p^0 x^0- \vec{p}_{k} \cdot \vec{x})}=0$. To solve them we need to 
recognize that the matrices in the chiral representation $S^{0i}$, $i=(1,2)$,
  transform ${\bf {\hat a}}^{\dagger}_{1}$ into ${\bf {\hat a}}^{\dagger}_{2}$,
  and opposite.},  for a plane wave 
$e^{i \varepsilon \vec{p}_{k} \cdot \vec{x}}$ for $|\vec{p_k}|=| p_k^0|$, 
then it follows 
%
\begin{eqnarray}
\label{relDirac}
\underline{\bf {\hat a}}^{h s \dagger}_{tot} (\vec{p}_k):=
(\alpha^{h s}_{\uparrow} (\vec{p}_{k}) \, {\hat a}^{h \dagger}_{\uparrow} + 
 \alpha^{h s}_{\downarrow} (\vec{p}_{k}) \,{\hat a}^{h \dagger}_{\downarrow}) \,
 e^{-i(p^0 x^0- \vec{p}_{k} \cdot \vec{x})} &=&
  \sum_{ m } {\hat {\bf a}}_{m }^{h \dagger} (\vec{p}_{k})
   v^{h s}_{m}(\vec{p}_{k})\, ,  
\end{eqnarray}
where the summation runs over $m $ up and down spin $m$
of the chosen handedness $h$.  

Since $v^{h s}_{m} (\vec{p}_{k})= \, u^{h s}_{m } (\vec{p}_{k})\,
e^{-i(p^0 x^0- \vec{p}_{k} \cdot \vec{x})} $ it follows also that 
${\bf {\hat a}}^{h s \dagger} 
(\vec{p}_k)= \sum_{m} u^{h s}_{m}\,  {\hat a}^{h \dagger}_{m} $, and 
$u^{h s}_{m} (\vec{p}_{k}) = \alpha^{h s}_{m} (\vec{p}_{k})$. We conclude
that $\underline{\bf {\hat a}}^{h s \dagger}_{tot} (\vec{p}_k)$ obviously determine 
${\hat {\bf a}}_{m}^{h \dagger} (\vec{p}_{k}) v^{h s}_{m}(\vec{p}_{k})=$
${\hat {\bf a}}_{m}^{h \dagger} (\vec{p}_{k}) \, u^{h s}_{m }(\vec{p}_{k})
e^{-i(p^0 x^0- \vec{p}_{k} \cdot \vec{x})}$.

Anticommutation relations of Eq.~(\ref{comDirac}), postulated by Dirac,
ensure the equivalent anticommutation relations also  for 
${\bf {\hat a}}^{h s \dagger} (\vec{p}_k )$ and 
${\bf {\hat a}}^{h s} (\vec{p}_k)$.

 Now we are able to relate creation and annihilation operators in both cases, the Dirac 
 case and our case of using the odd Clifford algebra to represent the internal space of fermions. 
\begin{eqnarray}
\label{DiracrelNorma}
\hat{b}^{h \dagger}_{1} &=& \stackrel{03}{[+i]}\stackrel{12}{(+)}\,\quad 
{\rm to\; be\; related\; to}\quad  {\hat a}^{h \dagger}_{\uparrow}\,,\nonumber\\
 \hat{b}^{h \dagger}_{2} &=& \stackrel{03}{(-i)}\stackrel{12}{[-]}\,\quad 
{\rm to\; be\; related\; to}\quad  {\hat a}^{h ^h\dagger}_{\downarrow}\,\nonumber\\
\hat{b}^h_{1} &=&-\stackrel{03}{[+i]}\stackrel{12}{(-)}\,\quad 
{\rm to\; be\; related\; to}\quad {\hat a}^h_{\uparrow}\,,\nonumber\\
 \hat{b}^h_{2}&=&\;\quad \stackrel{03}{(+i)}\stackrel{12}{[-]}\, \quad 
{\rm to\; be\; related\; to}\quad {\hat a}^h_{\downarrow}\,,
 \end{eqnarray}
 %
%
both sides representing the creation operators, with  $S^{12}= \frac{1}{2}$ and 
handedness $\Gamma^{(3+1)}=1$, Eq.~(\ref{hand}), in the first row,  and with 
$S^{12}= - \frac{1}{2}$ and handedness 
$\Gamma^{(3+1)}=1=h$, in the second row~%
\footnote{The vacuum state is on the left hand side  equal to 
$\stackrel{03}{[+i]}\stackrel{12}{[-]}$, while on the right hand side the 
corresponding vacuum state can be defined, if we follow our way of defining the 
vacuum state, to be proportional to $ ({\hat a}_{\uparrow}  
{\hat a}^{\dagger}_{\uparrow} + {\hat a}_{\downarrow}  {\hat a}^{\dagger}_{\downarrow})$.}.  

None of the creation operators, 
$ {\hat a}^{h \dagger}_{m}$, $m=(\uparrow, \downarrow)$ and 
$\hat{b}^{h \dagger}_{m}$,
$m=(1,2)$, depend on momenta, but ${\bf {\hat a}}^{h s \dagger} (\vec{p}_k)$
and ${\bf {\hat b}}^{ s f \dagger} (\vec{p}_k)$  as well as 
 $\underline{\bf {\hat a}}^{h s \dagger}_{tot} (\vec{p}_k)$
and $\underline{\bf {\hat b}}^{ s f \dagger}_{tot} (\vec{p}_k)$ do depend 
on momenta.





The creation operators $\underline{\hat{\bf a}}^{s \dagger}_{tot} (\vec{p}_k)$ 
fulfill the anticommutation relations of Eqs.~(\ref{Weylpp'comrel}, \ref{CliffcomrelHp}, \ref{ijthetaprodgenH}), the same as $\underline{\hat{\bf b}}^{s f \dagger}_{tot} (\vec{p})$ do.
We can just replace $\underline{\hat{\bf a}}^{s \dagger}_{tot} (\vec{p}_k)$ by 
$\underline{\hat{\bf b}}^{s f \dagger}_{tot} (\vec{p})$ for any of families (for plane waves solutions with continuous $\vec{p}$).

We can conclude:
\begin{eqnarray}
\label{DiracNorma}
\underline{\hat{\bf a}}_{tot}^{h s \dagger} (\vec{p}) 
&& \quad {\rm is\; to} \; {\rm be}  \; {\rm related}  \; {\rm to}\quad
\underline{\hat{\bf b}}^{h s  \dagger}_{tot} (\vec{p})\,,\nonumber\\
{\hat a}^{h\dagger}_{m}, m=(\uparrow,\downarrow) && \quad {\rm\;is \; to} \; {\rm be}  \; {\rm related}  \; {\rm to} \quad \hat{b}^{h \dagger}_{m}\, m=(1,2) \,,
\end {eqnarray}
with $h$ representing the handedness. This can be done for any chosen family in the Clifford case. In all the relations with  $\underline{\hat{\bf b}}^{h s  \dagger}_{tot} (\vec{p})$ the handedness is not written explicitly and is included in the index $m$ 
and in the index $s$, while the index $f$ represents the family quantum number.
Only in this chapter we introduce handedness in addition to clarify the relations.h

In the Clifford case the charges origin in spins $d\ge 6$. In $d=(13 +1)$
all the charges of quarks and leptons and antiquarks and antileptons can 
be explained, as well as the families of quarks and leptons and antiquarks 
and antileptons. In the Dirac case charges come from additional groups and so
 do families.

Let us add: The odd Clifford algebra influences the algebra  of the associated 
creation and annihilation operators acting on the second quantized Hilbert
 space ${\cal H}$; Due to oddness of the Clifford algebra, which determines 
 internal degrees of freedom of fermions, the creation operators and their
Hermitian conjugated  annihilation partners, determined on the tensor 
products of internal and momentum space, make the creation and annihilation
operators to anticommute.


We conclude:  The by Dirac postulated creation operators, 
$ {\hat {\bf a}}^{h \dagger}_{m} (\vec{p})$, and their annihilation partners, 
$ {\hat {\bf a}}^{h}_{m} (\vec{p})$, Eqs.~(\ref{stateDirac}, \ref{relDirac}), related 
in Eq.~(\ref{DiracNorma}) to the Clifford odd creation and annihilation operators, 
manifest that  the  odd Clifford algebra offers the explanation for  the second 
quantization postulates of Dirac.

\section{ Creation and annihilation operators in $d=(13 +1)$-dimensional space}
\label{13+1}
%

The {\it standard model} offered an elegant new step in 
understanding elementary fermion and boson fields by postulating:\\
{\bf i.} Massless family members of (coloured) quarks and (colourless) leptons, the left 
handed fermions as the weak charged doublets and the weak chargeless right hand 
members, the left handed quarks distinguishing in the hyper charge from the left 
handed leptons, each right handed member having a different hyper charge.
All fermion charges are in the fundamental representation of the corresponding groups. 
Antifermions carry the corresponding anticharges and opposite handedness. The  
massless families to  each family member exist.\\ 
 {\bf ii.} The existence of the massless vector gauge fields to the observed charges 
of quarks and leptons, carrying charges in the corresponding adjoint representations.\\
 {\bf iii.}  The existence of a massive scalar Higgs, gaining  at some step of the 
expanding universe the nonzero vacuum expectation value,  responsible for
masses of 
fermions and heavy bosons and for the Yukawa couplings. The Higgs carries a 
half integer weak charge and hyper charge.\\
{\bf iv.} Fermions and bosons are second quantized fields.

The {\it standard model} assumptions have in the literature several explanations, 
mostly with many new not explained assumptions. The most successful seem to be 
the grand unifying theories~\cite{Geor,FritzMin,PatiSal,GeorGlas,Cho,ChoFreu,Zee,%
SalStra,DaeSalStra,Mec,HorPalCraSch,Asaka,ChaSla,Jackiw,Ant,Ramond,Horawa,%
Kazakov2018},  if postulating in addition the family group and the corresponding 
gauge scalar fields.

The {\it spin-charge-family} theory, the project of one of the authors of this paper 
(N.S.M.B.~\cite{norma92,norma93,IARD2016,n2014matterantimatter,nd2017,%
n2012scalars,JMP2013,normaJMP2015,nh2017,nh2018}), is offering the explanation 
for all the assumptions of the {\it standard model}, unifying in 
$d=(13+1)$-dimensional space not only charges, but also charges and spins and 
families~\cite{norma93,nh02}, explaining the appearance of 
families~\cite{nh03,IARD2016,normaJMP2015}, the appearance of the vector 
gauge fields~\cite{JMP2013,nd2017}, of the scalar field and the Yukawa
 couplings~\cite{n2012scalars}. Theory offers the 
explanation for the dark matter~\cite{gn2009,gn2013}, for the matter-antimatter
asymmetry~\cite{n2014matterantimatter}, and makes several predictions~%
\cite{gn2014,gn2009,n2014matterantimatter}.

 The {\it spin-charge-family} theory is a kind of the Kaluza-Klein like 
theories~\cite{KaluzaKlein,Witten,Duff,App,SapTin,Wetterich,zelenaknjiga,mil,nh2017} 
due to the assumption that in $d\ge5$ (in the {\it spin-charge-family} theory 
$d\ge (13+1)$) fermions interact with the gravity only (vielbeins and two kinds of the
spin connection fields). Correspondingly this theory shares with the Kaluza-Klein 
like theories their weak points, at least: \\
{\bf a.} Not yet solved the quantization problem of the gravitational field.\\ 
{\bf b.} The spontaneous break of the starting symmetry, which would at low energies manifest the observed almost massless fermions~\cite{Witten}. \\ 
{\bf c.} The appearance of gravitational anomalies, what makes the theory not well defined~\cite{AlvarezWitten}, but in the low energy limit the fields manifest 
in $d=(3+1)$ properties of the observed vector and scalar gauge fields.\\
{\bf d.} And other problems.\\
In the {\it spin-charge-family} theory fermions interact in $d=(13+1)$ with the gravity  only: with the spin connections (the gauge fields of $S^{ab}$ and of $\tilde{S}^{ab}$) and vielbeins (the gauge fields of momenta), with fermions as a condensate present,
breaking the symmetry (and with no other gauge fields present), manifesting at low energies in $d=(3+1)$ as the ordinary gravity and all the observed vector gauge 
fields. \\
It is proven  in Refs.~\cite{NHD,ND012}, that one can have massless spinors 
even after breaking the starting symmetry. Ref.~\cite{nd2017} proves, 
that at low enough energies, after breaking the staring  symmetry, the two spin 
connections manifest in  $d=(3+1)$ as the observed vector gauge fields, as well 
as the scalar fields, which offer the explanation for the Higgs and the Yukawa 
couplings. Ref.~\cite{n2014matterantimatter} offers the explanation for the 
matter-antimatter asymmetry due to the existence of the scalar fields with the
``colour charges'' in the fundamental representations. In Ref.~\cite{nh2017} 
the {\it spin-charge-family} theory explains the {\it standard model} triangle anomaly cancellation better than the $SO(10)$ theory~\cite{FritzMin}. 

The working hypotheses of the authors of this paper (in particular of N.S.M.B.) 
is, since the higher dimensions used in the {\it spin-charge-family} theory 
offer in an elegant (simple) way explanations for the so many observed 
phenomena, that they should not be excluded by the renormalization 
and anomaly arguments. 
At least the low energy behavior of the spin connections and vielbeins as vector 
and scalar gauge fields manifest as the known and more or less well defined 
theories. 

In this  paper we present that using the half of the odd Clifford algebra objects 
to explain the internal degrees of freedom of fermions (the other half represent 
the Hermitian conjugated partners), as suggested by the 
{\it spin-charge-family} theory, leads to the second quantized fermions without
postulates of Dirac~\footnote{ 
The authors of Ref.~\cite{BPSN} let us know that their path integral formulation 
enabled them to see a great deal of what we present in this paper. We went 
through their paper noting that they did a lot concerning path integral formulation 
of quantum mechanics, offering ways to treat  anomalies. 
But we couldn't recognize that they propose some replacement for the Dirac 
postulates of creation and annihilation operators. We also could not found whether 
our proposal for explaining the Dirac postulates would bring any new light on 
path integral formulations and anomalies cancelations. To clarify this topics the discussions with authors would be needed.}.


%
\section{Conclusions}
\label{conclusionsCliff}

We present in Part I and Part II of this paper that the description of the internal 
space of fermions with the odd elements of the anticommuting algebra defines
the creation and annihilation operators, which anticommute when applied on 
the corresponding vacuum state. The internal space, described by the odd Clifford
algebra, extends its oddness to 
creation and annihilation operators generated on the tensor products of the 
internal basis with finite numbers of elements and the momentum basis
with infinite number of elements. The application of these creation and 
annihilation operators on the Hilbert space, determined by the tensor 
multiplication of all possible creation operators of any numbers, formally 
observed in this paper and in~\cite{nh2018} in the Clifford algebra, manifests 
the same anticommutation relations as the creation and annihilation of the 
second quantized fermions, explaining therefore the Dirac postulates of the 
second quantized fermion fields.

In the subsection~\ref{HNsubsection} we clarify the relation between our 
description of the internal space of fermions with ''basis vectors'', manifesting 
oddness and transferring the oddness to the corresponding creation and annihilation operators of second quantized  fermions, to the ordinary second quantized creation and annihilation operators from a  generalized  point of view.

We learn in Part I  of this paper, that odd products of superposition of $\theta^a$'s, 
Eqs.~(8-11,13,22) in Part I,
exist forming the odd algebra ''basis vectors'' in the internal space of ''Grassmann fermions'' with integer spin, which together with their Hermitian conjugated partners 
fulfill on the algebraic level on the vacuum state all the requirements for the anticommutation relations for the Dirac fermions. The creation and annihilation 
operators, defined on the tensor  products of the superposition of the Grassmann
odd algebra ''basis vectors'' and the momentum space basis, and manifesting 
correspondingly the oddness of the ''basis vectors'', fulfill the anticommutation relations of the second quantized Dirac's fermions on the vacuum state, as well 
as on the "Slater determinants'' of all possibilities of occupied and empty single particle ''Grassmann fermion'' states of integer spins of any number. These 
''Slater deerminants'', representing the Hilbert space of second quantized 
''Grassmann fermions'', can be represented as well with the tensor product multiplication of any possible choice of single ''Grasmann fermion states'' of all 
possible numbers of states, started with none (that is with the identity), 
distinguishing at least either in one of the quantum numbers of the ''basic 
vectors'' or in momentum basis.

%

In Part II we learn, that  the creation and annihilation operators  exist in the 
Clifford  odd algebra, defining the internal space of half integer fermions, which applying on  the vacuum state fulfill the anticommutation relations  postulated 
by Dirac. 
Creation operators, defined on the tensor products of the superposition of the 
finite ''basis vectors'' of the internal space described with the  Clifford algebra 
and of the infinite momentum basis, fulfill as well together with their 
Hermitian conjugated annihilation operators
the anticommutation relations  postulated by Dirac, on the vacuum state and  
on the Hilbert space of infinite number of the single particle fermion states,
$N_{\cal H}= \prod_{\vec{p}}^{\infty} 2^{2^{d-2}}$, 
Eqs.~(\ref{H}, \ref{NH}), creating  "Slater determinants" (Eqs.~(\ref{SD}, \ref{ijthetaprodgenH})), but only after the reduction of the degrees of 
freedom of the Clifford algebras for a factor of two, 
Eq.~(\ref{tildegammareduced}).

The reduction  of the Clifford algebras  for the factor of two leaves the 
anticommutation relations of Eqs.~(\ref{gammatildeantiher}, \ref{sabtildesab})
unchanged,  enabling  the appearance of family quantum numbers. 
The Clifford fermions carry half integer spins, families and charges in fundamental 
representations, Eq.~(\ref{eigencliffcartan}).


The reduction of Clifford space causes the reduction also in Grassmann space, what
leads to the disappearance of integer spin fermions, Eq.~(\ref{tildegammareduced1}).

The Clifford algebra oddness of the ''basis vectors'', describing the internal space of 
fermions, makes odd also the corresponding fermion states defined on the tensor 
products of the internal and momentum space. Correspondingly any two states 
fulfill the anticommutation relations and so do any  tensor products of odd numbers
of fermion states, forming the Hilbert second quantized space.

We present the creation operators, defined on the tensor products of 
''basis vectors'' and 
plane waves,  solve the equations of motion, in our case for free massless
 fermions, Eq.~(\ref{Weyl}), derived from the action, Eq.~(\ref{actionWeyl}).
 
Anticommutation relations are not postulated, as it is in the Dirac case, 
they follow from the oddness of the Clifford objects, and correspondingly explain
the second postulates of Dirac  (what is stressed in several places in Part I 
and Part II and in a short way also in Subsect.~\ref{HNsubsection}).

The relation between the Dirac's creation and annihilation operators and the 
ones offered by  the odd Clifford algebra,  discussed in In Subsect.~\ref{bethenormarelation}
demonstrates that the basic differences between these two descriptions is on the
level of the single particle creation operators: While the odd Clifford algebra 
offers the creation and annihilation operators, which fulfill the anticommutation 
relations, already on the level of the ''basis vectors'' determining the internal 
space of fermions, Eq.~(\ref{alphagammatildeprod}), when applied on the vacuum state, 
%
Dirac postulates the anticommutation relations  on the level of second quantized 
objects, following the procedure of Lagrange and Hamilton.

The final result is in both cases equivalent, leading to the Hilbert space of second
quantized fields. However, our way not only explains the Dirac postulates but 
demonstrates in addition, that also the single particle states in the first 
quantization do anticommute due to the oddness of the ''basis vectors'' 
defining the internal space. 
The oddness of the Clifford objects of creation and annihilation operators is 
transmitted from the ''basis vectors'' of internal space to the tensor products of 
the superposition of the  ''basis vectors'' and  the momentum or coordinate 
space. 


Correspondingly the odd Clifford algebra, equipped with the family quantum numbers,  
and fulfilling the anticommutation relations already on the level of the single particle 
creation operators applying on the vacuum state, as well as  on the level of  the whole 
Hilbert space, offers the explanation for the anticommutation relations, postulated by 
Dirac.




The Hilbert space of all ''Slater determinants'' 
with any number of occupied or empty  states of an odd character, follows in all 
three cases, the Dirac one (with postulated creation and annihilation operators and offering no families and no charges), the Grassmann one (offering spins and 
charges in adjoint representations, and no families) and the Clifford one (offering spins, families and charges),  in an equivalent way: due to the anticommuting 
creation and annihilation operators, representing ''basis vectors'' and their Hermitian conjugated partners. One can see  this in Sect.~\ref{bethenormarelation}.

Let us repeat: Internal space contributes the final number of states, the 
infinity of number of states is due to momentum/coordinate space~%
\footnote{Let us add that the single particle vacuum state is the sum of products
of annihilation $\times$ creation operators: In the Grassmann case it is just an identity, in the Clifford case is the sum of products of projectors for each family.}.


The anticommuting single fermion states manifest correspondingly the
oddness already on the level of the first quantization. 
Correspondingly these odd fermion states form in the tensor products
$*_{T}$ the Hilbert space {\cal H} of second quantized states.



%
\appendix 
\section{Norms in Grassmann space and Clifford space}
\label{normgrass}
%


Let us define the integral over the Grassmann  space~\cite{norma93} of two functions of the 
Grassmann coordinates $<{\cal {\bf B}}|\theta> <{\cal {\bf C}}|\theta>$, $<{\cal {\bf B}}| \theta>= 
<\theta | {\cal {\bf B}}>^{\dagger}$,  
\[<{\cal {\bf b}}|\theta>= \sum_{k=0}^{d} b_{a_1\dots a_k}
\theta^{a_1}\cdots \theta^{a_k},\] 
by requiring 
\begin{eqnarray}
\label{grassintegral}
\{ d\theta^a, \theta^b \}_{+} &=&0\,, \,\;\;  \int d\theta^a  =0\,,\,\;\; 
\int d\theta^a \theta^a =1\,,\;\; \nonumber\\
\int d^d &&\theta \,\,\theta^0 \theta^1 \cdots \theta^d =1\,,
\nonumber\\
d^d \theta &=&d \theta^d \dots d\theta^0\,,\,\;\; 
\omega = \Pi^{d}_{k=0}(\frac{\partial}{\;\,\partial \theta_k} + \theta^{k})\,,
\end{eqnarray}
with $ \frac{\partial}{\;\,\partial \theta_a} \theta^c = \eta^{ac}$. We shall use the weight function~%
\cite{norma93} 
$\omega= \Pi^{d}_{k=0}(\frac{\partial}{\;\,\partial \theta_k} + \theta^{k})$ to define the scalar
product  in Grassmann space $<{\cal {\bf B}}|{\cal {\bf C}} >$ 
\begin{eqnarray}
\label{grassnorm}
<{\cal {\bf B}}|{\cal {\bf C}} > &=&  \int 
 d^d \theta^a\, \,\omega 
 <{\cal {\bf B}}|\theta>\, <\theta|{\cal {\bf C}}> \nonumber\\
 &=& \sum^{d}_{k=0} 
\, b^{*}_{b_{1} \dots b_{k}} c_{b_1 \dots b_{k}}\,.%
\end{eqnarray}%
%

 To define norms in Clifford space Eq.~(\ref{grassintegral}) can be used as well.


%
\section{Handedness in Grassmann and Clifford space}
\label{handednessGrassCliff}

The handedness $\Gamma^{(d)}$ is one of the invariants of the group $SO(d)$, 
with the infinitesimal generators of the Lorentz group $S^{ab}$,
defined as 
\begin{eqnarray}
\label{handedness}
\Gamma^{(d)}&=&\alpha \varepsilon_{a_1 a_2\dots a_{d-1}} a_d\, S^{a_1 a_2} 
\cdot S^{a_3 a_4} \cdots S^{a_{d-1} a_d}\,,
\end{eqnarray}
with $\alpha$, which is chosen so that $\Gamma^{(d)}=\pm 1$.

In the Grassmann case  $S^{ab}$  is defined in Eq.~(\ref{sabtildesab}), while in the Clifford case
Eq.~(\ref{handedness}) simplifies, if we take into account that $S^{ab}|_{a\ne b}= 
\frac{i}{2}\gamma^a \gamma^b$  and $\tilde{S}^{ab}|_{a\ne b}= 
\frac{i}{2}\tilde{\gamma}^a \tilde{\gamma}^b$, as follows
\begin{eqnarray}
\Gamma^{(d)} :&=&(i)^{d/2}\; \;\prod_a \quad (\sqrt{\eta^{aa}} \gamma^a), 
\quad {\rm if } \quad d = 2n\,. 
\label{hand}
\end{eqnarray}
%


\begin{acknowledgements}
The author N.S.M.B. thanks Department of Physics, FMF, University of Ljubljana, Society of 
Mathematicians, Physicists and Astronomers of Slovenia,  for supporting the research on the 
{\it spin-charge-family} theory, the author H.B.N. thanks the Niels Bohr Institute for
being allowed to staying as emeritus, both authors thank DMFA and  Matja\v z Breskvar of 
Beyond Semiconductor for donations, in particular for sponsoring the annual workshops entitled 
"What comes beyond the standard models" at Bled. 
\end{acknowledgements}



\end{document}